\documentclass[%
 reprint,
superscriptaddress,
 amsmath,
 amssymb,
 aps,
 prl,
]{revtex4-2}

\usepackage{graphicx}
\usepackage{dcolumn}
\usepackage{bm}
\usepackage{hyperref}

\usepackage{xcolor}

\begin{document}

\title{Suppression of errors in collectively coded information}
\author{Martin J. Falk*}%
\affiliation{Department of Physics, The University of Chicago, Chicago, IL 60637}%
\author{Leon Zhou*}
\affiliation{Department of Physics, The University of Chicago, Chicago, IL 60637}%
\author{Yoshiya J. Matsubara}%
\affiliation{Department of Physics, The University of Chicago, Chicago, IL 60637}%
\author{Kabir Husain}%
\affiliation{Department of Physics and Astronomy, University College London, London WC1E 6BT, United Kingdom}
\affiliation{Laboratory for Molecular Cell Biology, University College London, London WC1E 6BT, United Kingdom}
\author{Jack W. Szostak}
\affiliation{Howard Hughes Medical Institute, Dept. of Chemistry, The University of Chicago, Chicago, IL 60637}%
\author{Arvind Murugan}
\affiliation{Department of Physics, The University of Chicago, Chicago, IL 60637}%

\date{\today}

\begin{abstract}
Modern life largely transmits genetic information from mother to daughter through the duplication of single physically intact molecules that encode information. However, copying an extended molecule requires complex copying machinery and high fidelity that scales with the genome size to avoid the error catastrophe. Here, we explore these fidelity requirements in an alternative architecture, the virtual circular genome, in which no one physical molecule encodes the full genetic information. Instead, information is encoded and transmitted in a collective of overlapping and interacting segments. Using a model experimental system of a complex mixture of DNA oligomers that can partly anneal and extend off each other, we find that mutant oligomers are suppressed relative to a model without collective encoding.  Through simulations and theory, we show that this suppression of mutants can be explained by competition for productive binding partners. As a consequence, information can be propagated robustly in a virtual circular genome even at mutation rates expected under prebiotic conditions.
\end{abstract}

\maketitle

Faithful copying of heritable information is a basic requirement for genomes. Standard accounts of fidelity emphasize enzyme-based mechanisms ranging from nucleotide selectivity\cite{echols1991fidelity} and exonucleolytic proofreading\cite{brutlag1972enzymatic} to post-replicative mismatch repair\cite{modrich1996mismatch} acting on a single, continuous template. These mechanisms are powerful but in extant biology are based on sophisticated protein machinery, and they must operate below the well-known error-catastrophe threshold to maintain information\cite{eigen1971selforganization}. A complementary approach, often overlooked, involves changing the architecture of the genome, i.e., how genetic information is physically laid out. Biology offers many precedents for diverse architectures: the fragmented chromosomes and plasmids of \emph{Borrelia burgdorferi}\cite{fraser1997genomic}; the linear, circular and branched forms of mitochondrial genomes\cite{chen2005organization}; the partitioning into a micronucleus and macronucleus in ciliates\cite{arnaiz2012paramecium}; and the interlocked kinetoplast DNA networks of many parasites\cite{lukeš2002kinetoplast}. Yet the impact of such architectural choices on how errors arise and propagate remains poorly understood.

\begin{figure*}
\centering
\includegraphics[width=6.2in]{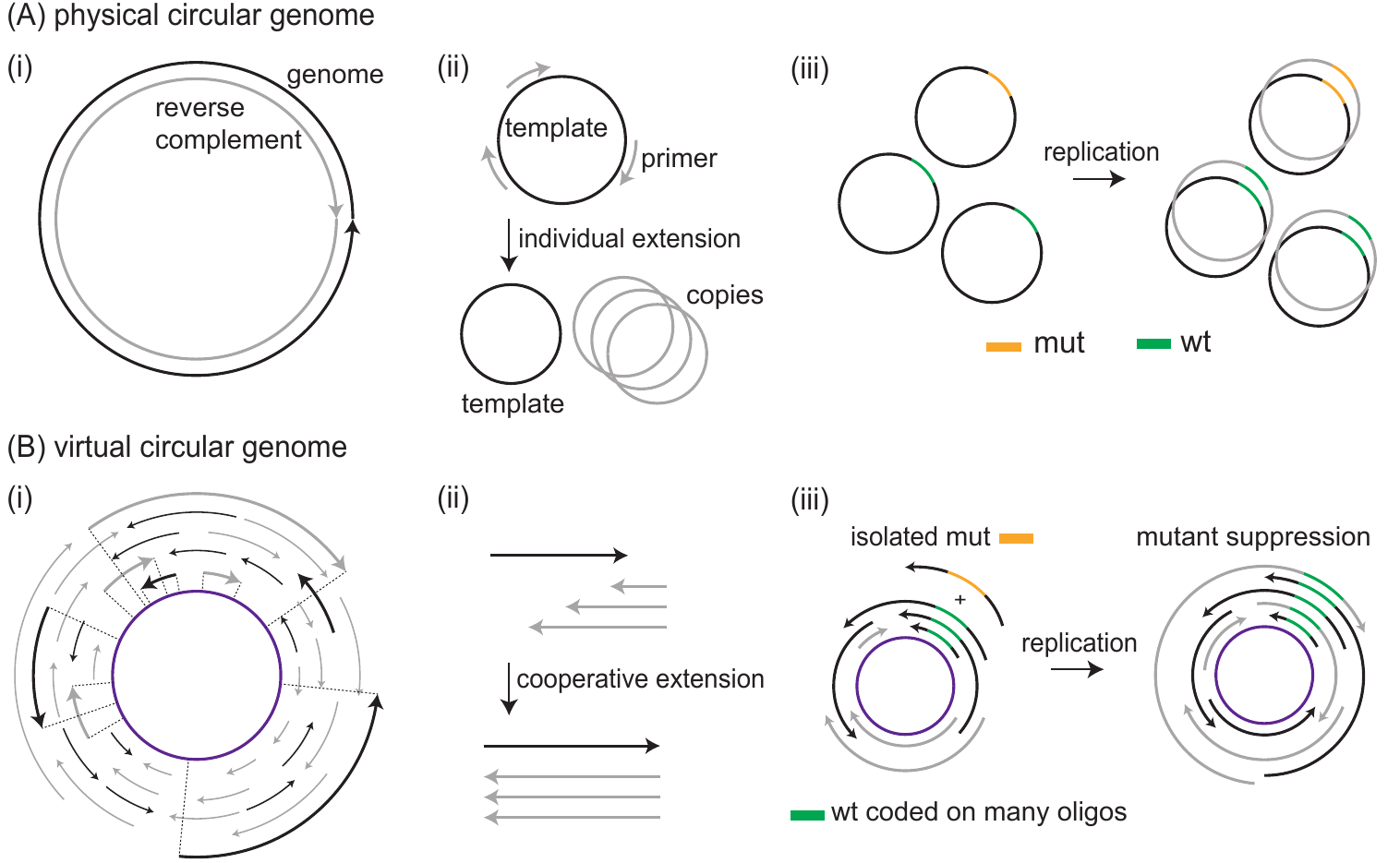}
\caption{Cooperative replication of genetic information in a virtual circular genome. (A) (i) In physically circular genomes found in extant life, genetic information is encoded in long nucleic acid polymers (ii) that are replicated by the extension of short primers. (iii) As each primer is extended to cover the entire genome, a neutral mutant allele in one part of the genome is replicated just as often as its wildtype counterpart. (B) (i) In contrast, the proposed virtual circular genome of a protocell is the consensus sequence (purple) of many short oligomers (black and grey). (ii) Each oligomer (oligo) may act as both primer and template during replication. (iii) Here, we show that the resulting co-operative effects, in which wildtype and mutant oligos compete as both primers and templates, suppress the replication of isolated mutant alleles in favor of wildtype alleles already coded on many oligos.}
\label{fig:intro}
\end{figure*}

Alternative architectures for genomic information are also motivated by the search for minimal self-replicating systems, whether envisioned for synthetic biology or as models of early life. Copying an extended template end-to-end typically requires elaborate machinery that can unwind and manage long duplex regions; in modern cells this role is played by the multi-enzyme replisome (helicases, polymerases, gyrases and many other accessory factors). Consequently, even engineered minimal cells currently rely on sizable gene sets to support these processes\cite{hutchison2016design}. Such machinery is unavailable in early evolutionary settings and in more minimal synthetic constructs. Hence it is natural to consider distributed architectures in which information is carried by many short genomic fragments and replication proceeds via many short, parallel extensions rather than a single long uninterrupted pass\cite{prywes2016nonenzymatic,mutschler2018random,derr2012prebiotically,toyabe2019cooperative,kudella2021structured,rosenberger2021self,calacca2024replication,mizuuchi2023minimal,winfree1998whiplash}. We refer to this class of architectures where a population of short, overlapping strands jointly stores and transmits a sequence as a collective encoding.

These considerations motivate the central question we address here: Do collectively encoded genomes, by virtue of their architecture alone, improve replication fidelity? We address this question in the context of the virtual circular genome (VCG) framework\cite{zhou2021virtual,ding2023experimental}, in which overlapping genomic fragments map to a circular consensus sequence, even though no single strand contains the whole. We use an experimental DNA-based implementation of a VCG, together with simulations and theory, to compare how wildtype and mutant sequences propagate in a collectively encoded pool.

Our results reveal an inherent asymmetry created by the architecture. Wildtype sequences that match the consensus are distributed across many different strands and hence retain many routes to continued copying. In contrast, mutations appear at singular locations and hence can have reduced routes to copying. The deficit is most severe for changes near the growing 3$^\prime$ end, where the location both cuts off continuation options and increases the likelihood of stalling—premature halts that maroon mutated pieces in short, non-contributing products—consistent with prior observations and models of stall-mediated error suppression and its possible role as a precursor to proofreading\cite{huang1992extension,ichida2005high,rajamani2010effect,goppel2021kinetic}. Consequently, mutants are intrinsically suppressed relative to wildtype, with the strongest effects for 3$^\prime$-terminal changes.

Our work shows that by distributing information across interacting oligomers, collective encoding provides an architecture-level route to fidelity that complements enzymatic error correction. This alternative is especially pertinent to early life forms, where elaborate enzymes may be absent, and to synthetic self-replication, where robust information transmission from simple parts is a design goal\cite{zhou2021mutations,ameta2021self,kriebisch2024template}.

\section{Experimental model} 
\label{sec:experimental_results}

\begin{figure*}
\includegraphics[width=.9\linewidth]{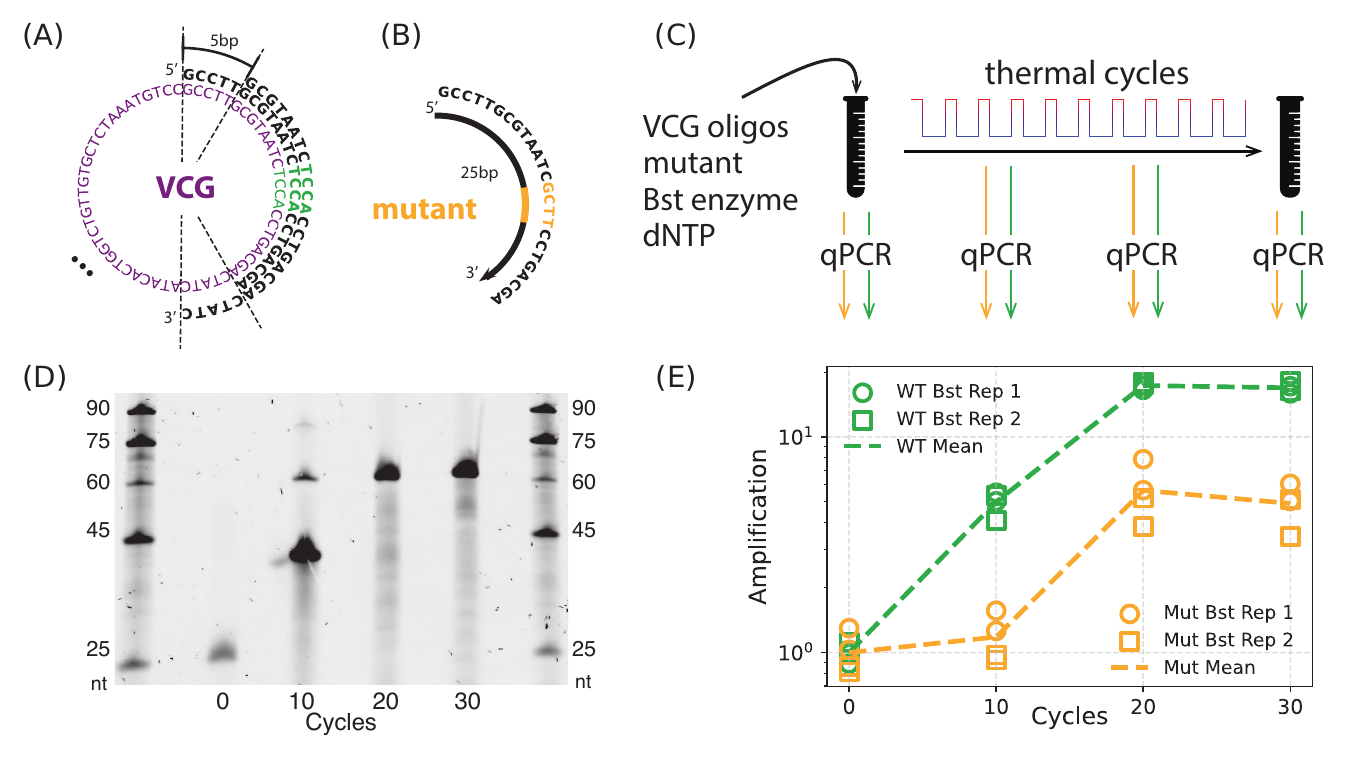}
 \caption{
 Suppression of mutant propagation in an experimental DNA model of the virtual circular genome (VCG). 
  (A) 60~bp consensus sequence for the VCG analyzed here;  sequence designed to avoid repeats of length $\geq$ 4~bp. VCG was synthesized as 24 DNA oligomers (12 each clockwise and counterclockwise), each one 25~nt in length and staggered 5~bp along the consensus sequence. 
 (B) Mutant oligos were designed by replacing a 4~bp region of one VCG oligo with a mutated sequence.
(C) Experimental setup for VCG replication. Mix of VCG and mutant oligos (here, initial mutant oligo concentration is 5\% of corresponding wildtype (wt) concentration) is combined with dNTPs (1~mM), Bst DNA polymerase, and 1$\times$ Bst buffer and subject to thermal cycling. 
(D) Denaturing gel electrophoresis of samples after 0, 10, 20, and 30 thermal cycles. DNA ladders (outermost lanes) range from 25 to 90~nt. The initial 25~nt oligos at cycle 0 progressively extend, with predominant products reaching $\sim 60$~nt by 20 cycles, consistent with the melting temperature $T_\mathrm{m} = 80~^\circ\mathrm{C}$ used in the cycles.
(E) Amplification of wildtype and mutant oligos as a function of thermal cycle, for 5\% initial mutant levels. Replication is quantified by qPCR every 10 thermal cycles, using primers specific to either a wildtype (green) or mutant (orange) allele sequence. Measured cycle threshold (Ct) values are converted to absolute DNA concentrations by a standard curve obtained by serial dilution of a known concentration. Amplification of an oligo is defined as oligo concentration normalized to its initial concentration before thermal cycling. Here, as later, circle and square markers denote two independent thermal-cycle replicates, each measured by qPCR in duplicate. 
}
\label{fig:experiment}
\end{figure*}

\begin{figure*}
\centering
\includegraphics[width=6.2in]{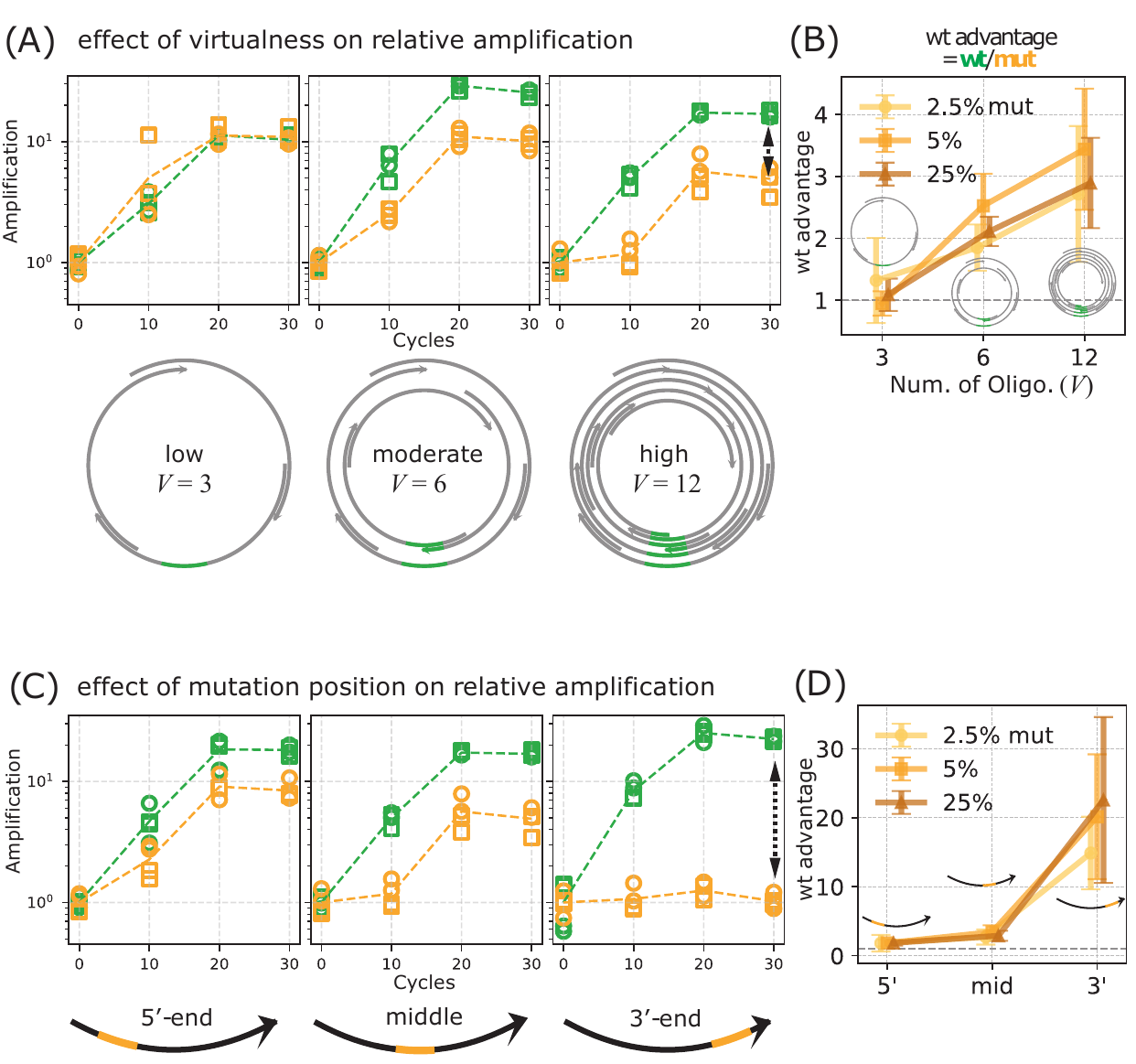}
\caption{
Mutant suppression depends on both the degree of `virtualness' of the virtual circular genome (VCG) and the proximity of the mutation to  3$^\prime$ end of the oligomer. 
(A, B) Effect of genome virtualness $V$. Initial VCG pools were constructed from 3, 6, or 12 overlapping oligos in each direction (plus their reverse complements), representing increasing levels of virtualness. 
(A) Amplification of wt and mutant oligos as a function of thermal cycle at each level of virtualness, shown as fold-change relative to the initial concentration (y-axis) across thermal cycles (x-axis). Concentrations were inferred from qPCR Ct values, as in Fig.~\ref{fig:experiment}E. The initial concentration of mutants was set at 5\% of the VCG oligos.
(B) wt advantage as a function of virtualness $V$.  wt advantage is the ratio of wt to mutant amplification at cycle 30. Different lines represent results from different initial concentrations of mutants. wt advantage is strongest in highly virtual VCGs. In contrast, wt advantage is lost when virtualness is low, approximating a physical circular genome.
(C, D) Effect of mutation position. Oligos contain a mutant region at varying positions along the sequence. 
(C) Amplification of wildtype (wt) and mutant oligos.   
(D) Amplification advantage of wt over mutant (wt advantage) plotted against the mutation position on the mutant. wt advantage is strongest when the mutation is located at the 3$^\prime$ end. }
\label{fig:virtualness}
\end{figure*}

We designed a DNA-based system composed of short, overlapping oligomers (oligos) that form a redundant circular architecture. The full VCG comprises 12 distinct 25~nt oligos and their reverse complements. Each of the 12 oligos in one direction overlaps with its downstream neighbor (i.e., the next oligo in the 5$^\prime$--3$^\prime$ direction) by 20 bases,  encoding a 60~bp circular genome (Fig.~\ref{fig:experiment}A, Supp.~Table~I). Thus, each oligo initially has four \emph{binding partners} that productively allow it to be extended or part of it to be copied, i.e.,  reverse-complementary oligos that can hybridize and serve as templates and/or as primers,  with partial complementarity ranging from 5 to 20~bp. Each oligo also has an additional five binding partners which anneal in unproductive configurations that do not allow for primer extension. Less structured DNA experimental models with pools of oligos have been investigated in prior work\cite{kudella2021structured,calacca2024replication,luther1998surface,sievers1994oligomers,derr2012prebiotically,rosenberger2021self,toyabe2019cooperative}.

To evaluate how mutations propagate within this architecture, we introduced a 25~nt mutant oligo that differs from one VCG oligo by a single 4~bp substitution near the center of the sequence (Fig.~\ref{fig:experiment}B, Supp.~Sec.~I). Mutant oligos were spiked into VCG mixtures at defined low proportions (2.5\%, 5\%, or 25\% relative to the wildtype) to mimic rare variant emergence. 

Extension of VCG oligos was driven by thermal cycling with Bst DNA Polymerase (Large Fragment, NEB M0275), using 10--30 cycles of alternating denaturation ($80~^\circ\mathrm{C}$) and annealing/extension ($35~^\circ\mathrm{C}$), followed by a final enzyme deactivation step ($90~^\circ\mathrm{C}$) for 10 minutes. Notably, {no new oligomers are supplied to this system}; extension is initiated by overlap-driven annealing of partially complementary VCG oligos (Supp.~Sec.~III). 
Gel electrophoresis confirmed that initial 25~nt VCG oligos were extended incrementally over cycles, {with products reaching a length of $\sim$45~nt by cycle 10}, and further reaching $\sim$60~nt by cycle 20, with gel profiles remaining unchanged through cycle 30 (Fig.~\ref{fig:experiment}D). Duplexes up to 45~bp ($T_\mathrm{m} = 80~^\circ\mathrm{C}$) likely remain partially meltable during cycling, allowing oligo reshuffling and continued extension, whereas ~60~bp duplexes ($T_\mathrm{m} = 82~^\circ\mathrm{C}$) are too stable to denature, halting further growth. We refer to this final state as `pool stasis'.

To quantify the relative amplification of wildtype and mutant alleles, we used sequence-specific qPCR with forward primers that selectively bind either the wildtype or mutant sequence, combined with a shared reverse primer (Fig.~\ref{fig:experiment}C, Supp.~Sec.~IV). Two aliquots from each thermocycled sample were analyzed in parallel using both primer sets, allowing independent quantification of wildtype and mutant allele concentrations (Fig.~\ref{fig:experiment}D–E). Ct values were determined from the qPCR amplification curves and converted into absolute concentrations using standard curves generated from known DNA concentrations included in the same qPCR run. Control experiments confirmed that the qPCR assay reliably quantifies each allele sequence with high specificity and within the relevant concentration range (Supp.~Fig.~S1).

qPCR analysis revealed that the wildtype allele was consistently amplified more than the mutant allele over 30 cycles of VCG extension; 
competitive oligo interactions favor wildtype proliferation and effectively suppress the mutant (Fig.~\ref{fig:experiment}E, Supp.~Sec.~VI). Notably, this suppression effect is time-dependent: before cycle 10 the mutant is strongly suppressed, while from cycles 10 to 20, wildtype and mutant grow with equal speed until pool stasis. We define a `wt advantage' metric as the ratio of wildtype and mutant amplification at pool stasis (i.e., at cycle 30).

To investigate the role of VCG architecture in mutation suppression, we varied the system's `virtualness' $V$ -- defined as the number of overlapping oligo pairs encoding the genome -- by constructing VCGs with 12, 6, or 3 oligo pairs (Fig.~\ref{fig:virtualness}A, Supp.~Table~II). 
All of these sets of oligos encode the same 60~bp genome but differ in the number of distinct potential reverse complement binding partners available to each oligo.
Lower virtualness (e.g., the 3-oligo VCG) more closely resembles a real physical circular genome in that it has low redundancy measured by how many different oligos cover a given subsequence of the genome. We observed stronger suppression of mutant amplification in higher-virtualness VCGs across all mutant input levels we tested,  including 2.5\% (Supp.~Fig.~S4), 5\% (Fig.~\ref{fig:virtualness}A; Supp.~Fig.~S5), and 25\% (Supp.~Fig.~S6).
In the 12-oligo VCG, the wildtype outcompeted the mutant by over 10-fold, while in the 3-oligo VCG, the amplification advantage for the wildtype was nearly absent (Fig.~\ref{fig:virtualness}B).

We next examined how the position of the 4~bp mutation within the mutant oligo affects its suppression. Using additional mutant variants with mutations placed near the 3$^\prime$ or 5$^\prime$ ends, we found that mutations near the 3$^\prime$ end were most strongly suppressed during extension (Fig.~\ref{fig:virtualness}C--D). In contrast, the 5$^\prime$ end mutant exhibited amplification levels nearly indistinguishable from the wildtype. This trend held consistently across all VCG virtualness levels and initial mutant proportions (Fig.~\ref{fig:virtualness}D, Supp.~Fig.~S4--6), suggesting that suppression is more effective when the mutation is closer to the 3$^\prime$ end---the site where oligo extension initiates. This positional bias aligns with the expectation that errors or mismatches near the 3$^\prime$ end {would exhibit a decreased likelihood of propagation}.

Together, these findings demonstrate that both VCG virtualness and mutation position govern selective replication within this oligo-extension-based system. 
The results suggest that VCG architectures can impose intrinsic fidelity constraints, naturally suppressing the propagation of mutant alleles relative to the wildtype allele during thermal cycles.

\section{Simulation}

\subsection{Simplified VCG Model}

\begin{figure*}
\centering
\includegraphics[width=.9\linewidth]{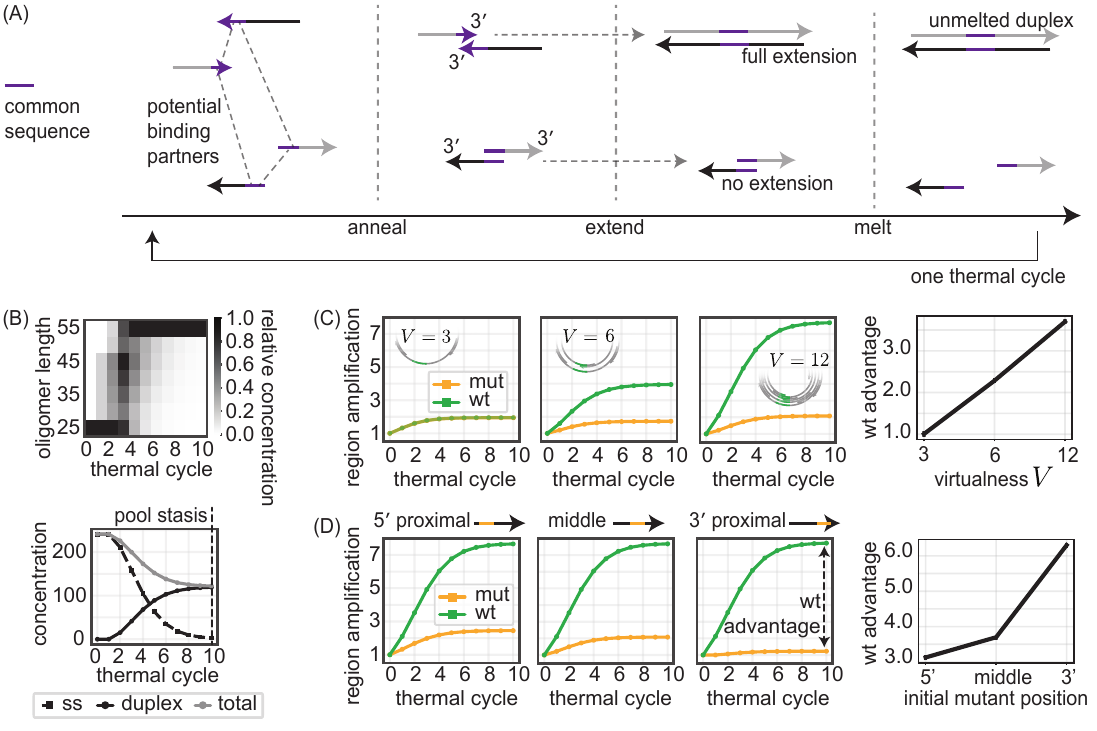}
\caption{Cooperative effects in a VCG simulation reproduce differential suppression of mutant alleles.
(A) Our (differential equations-based) simplified VCG model consists of thermal cycles with the following events: annealing, extension, and melting. Annealing: oligos with common sequence overlap $o$ above a minimum length ($o_{min} = 2$) can bind to each other to become duplexes. Extension: oligos with 3$^\prime$ ends annealed to their duplex templates extend as primers off of each other to form blunt ends. Melting: Duplexes with fewer than a maximum number ($o_{max} = 55$) of paired bases are allowed to melt. Initial oligo pools are chosen to match experimental oligo pools. See Supp.~Sec.~VII for further details. 
(B) (top) Distribution of oligo lengths after the melting step of each of 10 simulated thermal cycles. Relative concentration across each column is normalized by max binned concentration value.
(bottom) Concentration of single-stranded (ss) oligos, concentration of unmelted duplexes, and total concentration of all oligos after the melting step of each thermal cycle. `pool stasis' in simulations at cycle 10 indicates all oligos are bound in unmeltable long duplexes ($o \ge o_{max}$). 
(C) (left) mut and wt region amplification as a function of thermal cycle for three different values of VCG virtualness $V$, defined as number of distinct oligo pairs that make up initial oligo pool (either 3, 6, or 12). Region amplification is defined as the instantaneous concentration of an allele plus a flanking region (set by the oligo the mut allele initially appears on), normalized by region's initial concentration. (right) wt advantage (wt region amplification divided by mut region amplification at pool stasis, here 10 cycles) as a function of $V$. mut allele introduced in the middle of a single oligo for all $V$ conditions.
(D) (left) mut and wt region amplification for three different initial mut positions (5$^\prime$-proximal to 3$^\prime$-proximal) as a function of thermal cycle. (right) wt advantage as a function of initial mut position. {$V=12$ across different mut position conditions.}}
\label{fig:simulation}
\end{figure*}

To understand the source of position- and virtualness-dependent mutation fates in the VCG, we build a simplified model of VCG replication (Fig.~\ref{fig:simulation}A). In this model, we consider an initial pool of oligos which grows through repeated thermal cycles. 
This initial oligo pool is based on the experimental oligo pools studied in the previous section. 
It contains 12 oligo sequences of length 25~nt which form a consensus VCG sequence of length 60 bp with an offset of 5 bp between each consecutive oligo. In addition, the pool also contains all the reverse complements of the original 12 oligos, plus a small concentration of an additional oligo that is identical to the underlying VCG sequence except at a single position, where it contains a mutant allele.
This is in contrast to the experimental mutant oligo, which differs from its wt counterpart in a 4~nt block.

The initial pool undergoes thermal cycles which consist of the following:
\begin{enumerate}
    \item First, oligos anneal with any possible complementary partner to form duplexes, following irreversible second-order reaction kinetics until no further annealing reactions are possible. Only oligos which have exact contiguous overlaps above a threshold $o_{min}$ amount are allowed to bind. Here $o_{min} = 2$.
    \item {Second, if an oligo has an annealed 3$^\prime$ end, it acts as a primer and is extended using its annealed duplex partner as a template.} This extension continues without any errors until the extending oligo reaches either the 5$^\prime$ end of its template or a maximal length $l_{max}$ equal to the length of the consensus VCG sequence (60~nt).
    \item Finally, during the melting stage, duplexes dissociate into single-stranded oligomers unless their overlap exceeds a specified threshold length $o_{max}$. Here $o_{max} = 55$, as approximately observed in experiments.
\end{enumerate}

We note that annealing in any single thermal cycle is random in that the annealed duplexes are kinetically determined. Thus, only a small fraction of an oligo may be bound to its perfect complementary pair. 
For instance, in the initial oligo pool where oligos are of length 25~nt and are separated by offsets of 5~nt, then any oligo has $9$ potential binding partners, only one of which is its perfect complementary pair.
Binding the perfect complementary pair forms a blunt-end duplex which does not allow for further extension.
However, each oligo has an additional 8 other binding partners which can bind in equal probability; 4 of those 8 bind in duplexes which allow for extension.

Therefore, because this system anneals through irreversible kinetics, not all pairs bind with perfect complementarity and hence extension can occur through the formation of duplexes with overhangs. 
The formation of duplexes with perfect complementarity is further disrupted by the melting phase of thermal cycles on a timescale  $\tau_{cycle}$.
For more details on how our model is defined and implemented, please see Supp.~Sec.~VII.

\subsection{Mutant suppression in the simplified model}
Following simulation of repeated thermal melt-anneal-extend cycles, we first observe that oligos extend on each other (Fig.~\ref{fig:simulation}B, top); {the distribution of oligo lengths shifts from their initial values (25~nt) to a maximal length set by a combination of $o_{max}$ (the maximal duplex overlap length for unbinding during the melt phase) and $l_{max}$ (the maximal oligo length).} As a consequence of their extension, oligos become longer and more likely to be trapped in duplexes that overlap too much to unbind during the melting step. 
In particular, while oligos at the end of thermal cycle 0 are still all single-stranded, virtually all oligos are bound in duplexes of length 55--60~bp at the end of cycle 10 (Fig.~\ref{fig:simulation}B, bottom). We refer to this state where all oligos are duplex-bound as `pool stasis', since the VCG cannot replicate anymore. In simulations, we see pool stasis occurs at approximately cycle 10.

Extension also allows for an increase in the concentration of mut and wt alleles, despite the fact that no new oligos are created in our model. As oligos extend off each other as templates, new strands bind downstream of the mut and wt alleles (i.e., between the allele on an oligo and the 3$^\prime$ end of that oligo) and extend past those alleles, thereby increasing the concentration of oligos which contain the two alleles.

To model the qPCR readouts of the DNA model experiments, we track the relative amplification of the mut and wt alleles by measuring the total concentration of oligos that contain the allele along with a flanking region around the allele. This flanking region is set by the oligo that the mut allele initially appears on. After normalizing these concentrations to the initial concentration of flanking region oligos, we can define a `wt advantage' statistic as the ratio between the wt region amplification at pool stasis and the mut region amplification at pool stasis.
We find that the wt advantage increases with the virtualness $V$ of the initial oligo pool (Fig.~\ref{fig:simulation}C), where $V$ here can be defined as the number of unique oligo pairs that cover the consensus VCG sequence in the oligo pool (either 3, 6, or 12). We also find that the wt advantage measured at pool stasis is stronger as the initial mutant position shifts from the 5$^\prime$ end to the 3$^\prime$ of the oligo it starts on (Fig.~\ref{fig:simulation}D). 

\subsection{Collective binding partner effects predict mutant suppression}

\begin{figure*}
\centering
\includegraphics[width=.9\linewidth]{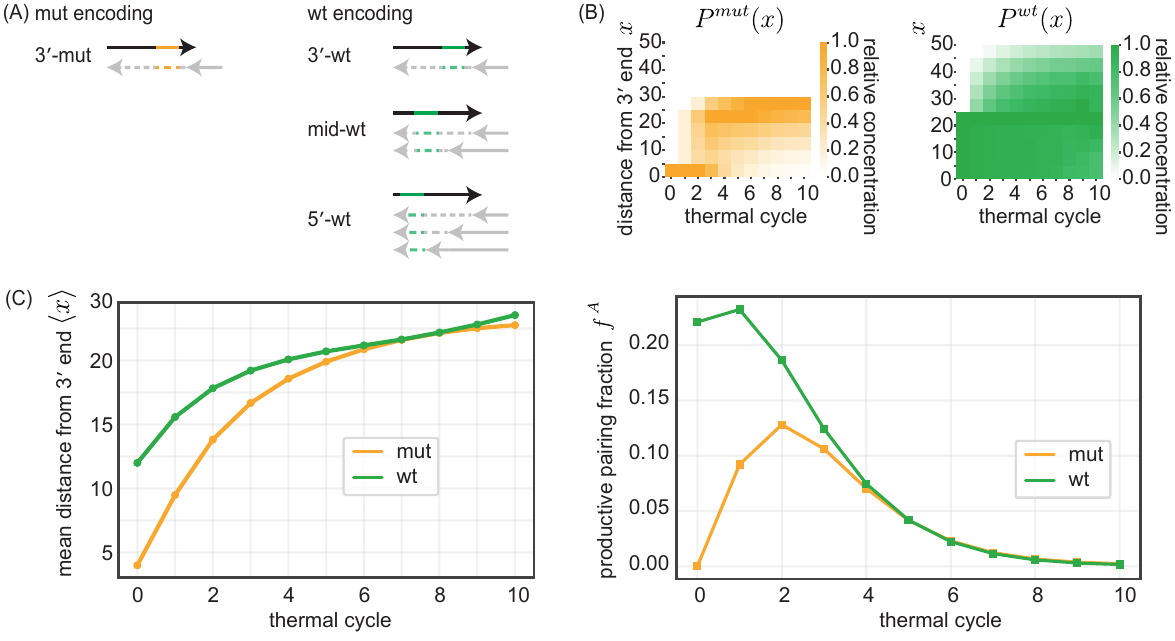}
\caption{A binding partner effect explains the suppression of initially 3$^\prime$-proximal mutants.
(A) A mutant allele located towards the 3$^\prime$ end of an oligo has only 1 `productive' binding partner capable of copying that mutant allele through extension (left schematic). In contrast, the wt allele is present at varying positions on different oligos, including 5$^\prime$ end placements that have many productive binding partners capable of copying it (right schematic). (B) Evolution of distributions for mut $P^{mut}(x)$ and wt $P^{wt}(x)$ allele distance $x$ from 3$^\prime$ end of oligos that contain the allele over thermal cycles. Distribution is for single-stranded oligos only. Relative concentration across each column is normalized by max binned concentration value.
(C) (left) Average distance of mut and wt allele from 3$^\prime$ end of oligos that contain allele as a function of thermal cycle.
(right) Evolution of the average productive pairing fraction $f^A$ for mutant and wt alleles as a function of thermal cycle. For oligo $i$ containing allele $A$, we define a productivity factor $F^A_i$ as the ratio between the concentration of single-stranded oligos that can bind to oligo $i$ downstream of $A$ and the total concentration of oligos which can bind to oligo $i$. For oligos only bound in duplexes, $F^A_i = 0$. The productive pairing fraction $f^A$ is a concentration-weighted average over $F^A_i$. 
}
\label{fig:explanation}
\end{figure*}

In order to find an explanation for the observation of mutant suppression both in our experimental DNA system as well as our simplified simulation model, we focus on the case where the mutant allele is initially at the 3$^\prime$ end of the oligo on which it occurs. This is both the situation where the mutant suppression is the most dramatic (Fig.~\ref{fig:simulation}D), as well as the most relevant for mutant alleles that arise through mistakes during extension. 

We propose that the wt advantage comes from a binding partner effect. In the VCG pool, each oligomer has multiple potential reverse complement binding partners, with the number of potential partners increasing with the virtualness $V = 3, \ldots, 12$ of the initial VCG oligo pool. These binding partners all compete for annealing with one another on the template oligo, on equal footing with the perfect reverse complement partner.

Crucially, only binding partners whose 3$^\prime$ ends anneal downstream of the allele (i.e., between the allele and the 3$^\prime$ end of the template), can increase that allele’s reverse complement concentration via extension. 
In contrast, binding partners whose 3$^\prime$ ends are upstream of the allele may extend but do not result in copying of the allele into its reverse complement, precluding an increase in concentration of the allele itself upon further rounds of extension.

Therefore, a mut allele initially located near the 3$^\prime$ end of an oligo has fewer binding partners that can bind downstream of that allele and make its reverse complement during extension. 
In contrast, wt alleles are initially distributed across many different oligo locations. When the wt allele is located towards the 5$^\prime$ end of a template oligo, it can be copied by almost all binding partners that anneal to that template (see Fig.~\ref{fig:explanation}A).

The binding partner effect implies that the instantaneous 
per-cycle rate at which an allele on a given oligo can produce more copies of its reverse complement is set by two quantities: 1.~the concentration of `productive' binding partner oligos which can bind and extend past the allele and are thus productive in copying that allele, and; 2.~the concentration of all binding partner oligos that compete with (1), regardless of whether they can pick up the allele upon extension. 
An allele is more productive if it exists further from the 3$^\prime$ oligo end and hence more of the oligo's binding partners can form pairings that allow the allele to be copied.
Correspondingly, an allele is also more productive if it sits closer to the 5$^\prime$ oligo end since there are fewer binding partners that do not pick up the allele upon extension. 

To formalize this intuition, we can define a productivity factor $F^A_i$ for an allele $A$ on a single-stranded oligo $i$ as:
\begin{align}
\label{eqn:oligo_productivity_factor}
    F^A_i \equiv \frac{\sum_{j \in D^A_i}c_{ss,j}}{\sum_{j \in B_i}c_{ss,j}},
\end{align}
where $c_{ss,j}$ is the concentration of single-stranded oligo $j$,  $D^A_i$ is the set of oligos that bind with 3$^\prime$ ends downstream of the allele $A$ on oligo $i$, {and $B_i$ is the set of all oligos that can bind to oligo $i$ regardless of the configuration of the duplex they form.} (When oligo $i$ only exists in duplexes, $F^A_i$ is defined to be 0.)

Given this per-oligo definition of allele productivity, we can go further and integrate productivity across all oligos that carry that allele. Specifically, to compute the rate at which an allele produces its reverse complement per-cycle 
and per-oligo, we take a concentration-weighted average of the individual oligo productivity factors $F^A_i$ and define a productive pairing fraction $f^A$:
\begin{align}
\label{eqn:allele_productivity_factor}
    f^A = \frac{1}{c^A_{tot}}\sum_{i \in O_A}c_{ss,i} ~F^A_i,
\end{align}
where $O_A$ is the set of oligos that contain the allele $A$ and $c^A_{tot}$ is the total concentration of oligos containing $A$ regardless if they are single-stranded or in a duplex. 

This definition captures the intuition of how the position of an allele on an oligo influences the allele's productivity. For instance, when oligos are shorter than the oligo length maximum $l_{max}$, an allele $A$ coded on the 5$^\prime$ end of oligo $i$ will have a maximal productivity factor $F^A_i = 1$. This high productivity comes because every binding partner of oligo $i$ is capable of copying $A$ through extension, so every binding partner oligo index $j \in D^A_i$ is also $j \in B_i$.
In contrast, $F^A_i = 0$ for that same allele $A$ if it were coded at the 3$^\prime$ end of oligo $i$; in this case, there are no downstream binding partners for $A$ on $i$, so $D^A_i$ is empty. 
An approximation of the link between allele position and productivity that interpolates between the 3$^\prime$ and 5$^\prime$ end location limits is therefore:
\begin{align}
F^A_i \sim x^A_i/l_i,
\end{align}
where $x^A_i$ is the distance of allele $A$ from the 3$^\prime$ end of oligo $i$, and $l_i$ is the length of oligo $i$; more generally, $F^A_i$ is a monotonically increasing function of $x^A_i$ which is 0 at $x^A_i = 0$.

Therefore, in order to understand the time evolution of $f^A$ for the wt and mut alleles, we should track the distributions $P^{mut}(x)$ and $P^{wt}(x)$ of an allele's distance $x$ from the 3$^\prime$ end of oligos across single-stranded oligos in the oligo pool (Fig.~\ref{fig:explanation}B). 
While the mutant allele is initially 3$^\prime$-proximal (concentrated at 5~nt from the 3$^\prime$ end), by cycle 10 $P^{mut}(x)$ eventually spreads out. 
$P_{wt}(x)$, in contrast, is by construction initially more spread out since the wt allele was already encoded on a variety of different oligos (e.g. see Fig.~\ref{fig:experiment}A), and remains uniformly spread after all 10 cycles. 
Tracking the average distance $ \langle x \rangle$ of the mut and wt alleles from the 3$^\prime$ end (Fig.~\ref{fig:explanation}C, left) reveals that the mutant allele is initially closer to the 3$^\prime$ end of oligos compared to the average wt allele, indicating an initial disadvantage in replicative potential. However, as the oligo pool replicates, the relative positions of the mean wt and mean mutant allele converge. 

Indeed, if we calculate the time-dependent productive pairing fraction $f^A$ (Fig.~\ref{fig:explanation}C, right), we find that mut alleles are initially at a disadvantage, with a lower productive pairing fraction $f^{mut}$ compared to wt allele $f^{wt}$. This disadvantage {decreases} over the first few replication cycles, consistent with the decreasing gap between mean wt and mut distances from oligo 3$^\prime$ ends. By cycle 5, the productive pairing fraction $f^A$ of wt and mut alleles are essentially equal and both very close to 0. The collapse of the productive pairing fraction $f^A$ for both wt and mut alleles is due to the fact that most alleles are bound in fully-extended blunt end duplexes. Such duplexes cannot participate in subsequent rounds of melt-anneal-extend and hence cannot contribute to the production of either wt or mut alleles.

In summary, the initial disadvantage faced by 3$^\prime$-proximal mutant alleles due to the binding partner effect is transient. However, the cumulative impact of this early disadvantage leads to a persistent suppression of the mutant allele relative to the wildtype, even by the time of pool stasis. We next discuss how the stalling effect\cite{ichida2005high,perrino1989differential,mendelman1990base, huang1992extension,johnson2004structures, goppel2021kinetic, Laurent2024-dk,rajamani2010effect,ravasio2024minimal,matsubara2023avoidance} can significantly prolong this transient phase, thereby amplifying the overall suppression of the mutant allele.

\section{Stalling enhances collective suppression of mutants}

\begin{figure}
\centering
\includegraphics[width=\linewidth]{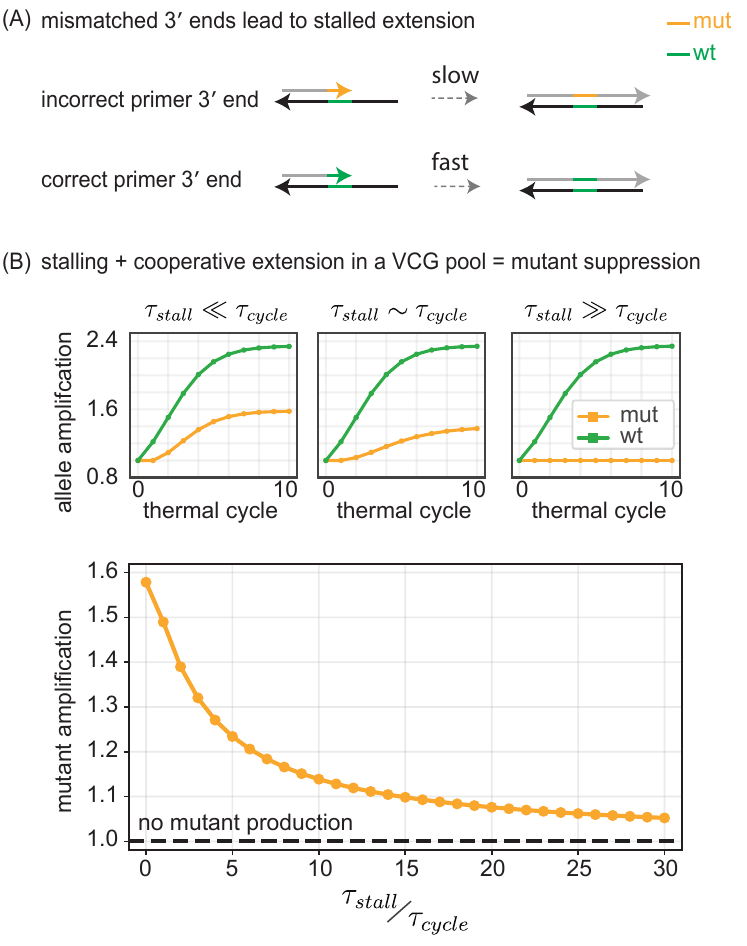}
\caption{Combination of collective effects and stalling suppresses mutant growth. (A) In duplexes where priming oligo has a correctly matched 3$^\prime$ end, extension proceeds normally and all such primers fully extend to form blunt ends. In duplexes where priming oligo has an incorrectly matched 3$^\prime$ end, extension past the mismatch is slowed by an additional time, $\tau_{stall}$. Consequently, only a fraction $1-e^{-\frac{\tau_{cycle}}{\tau_{stall}}}$ of mismatched oligos manage to extend within a thermal cycle of time $\tau_{cycle}$. See Supp.~Sec.~VII for further details.
(B) (top) mut and wt amplification for three different values of $\frac{\tau_{stall}}{\tau_{cycle}}$ as a function of thermal cycle. (bottom) Mutant allele amplification measured after 10 thermal cycles as a function of $\frac{\tau_{stall}}{\tau_{cycle}}$. }
\label{fig:stalling}
\end{figure}
The stalling effect describes the reduced rate of extension from a primer that ends in a mismatched base pair, compared to extension from a perfectly matched primer. 
Here, we investigate how stalling influences the propagation of mutant alleles within the VCG framework.

We consider annealed configurations where one of the oligos in the duplex is not correctly matched to its template at its 3$^\prime$ end. This can either happen when an oligo with a mutant allele at its 3$^\prime$ end is annealed to an oligo with a wt allele at that position, or it is an oligo with a wt allele at its 3$^\prime$ end annealed to an oligo with a mutant allele there. 
For such duplexes where one of the oligos has a mismatch at its 3$^\prime$ end, stalling delays extension by a time $\tau_{stall}$. Whether or not the duplex gets to extend therefore depends crucially on whether thermal cycling (with timescale $\tau_{cycle}$) drives the duplex apart before the stalling delay can be overcome. For primers with no 3$^\prime$ mismatch, there is no barrier to extension and such primers will fully extend during the current thermal cycle. To model the effect of stalling, we posit that primers with a mismatched 3$^\prime$ end get to extend through an exponential decay process with timescale $\tau_{stall}$. Hence by the end of a melt-anneal-extend cycle of length $\tau_{cycle}$, only a fraction $1-e^{-\frac{\tau_{cycle}}{\tau_{stall}}}$ of primers with a mismatched 3$^\prime$ end will have extended (Fig.~\ref{fig:stalling}A). Please see Supp. Sec. VII for further discussion and implementation details on stalled extension.

Stalling can have a dramatic effect on the suppression of mutant alleles in a VCG scenario depending on the relative balance of $\tau_{stall}$ and $\tau_{cycle}$ (Fig.~\ref{fig:stalling}B, top). When $\tau_{stall} \ll \tau_{cycle}$, we find the expected mutant suppression described earlier, which is due to the transient replicative ability reduction arising from the binding partner effect.
For intermediate $\tau_{stall} \sim \tau_{cycle}$, that transient period is greatly increased for the mutant because only a fraction of mismatched primers extends in any cycle; as a consequence, the mutant allele stays encoded at the 3$^\prime$ end for many more cycles and has a reduction in replicative ability. When $\tau_{stall} \gg \tau_{cycle}$, the mutation remains confined to the 3$^\prime$ end and there is no increase in mutant allele concentration at all.
Thus stalling provides a suppression of mutant alleles by lengthening the transient period of reduced replicative ability for the mutant. We calculated the amplification of mutant alleles at the end of 10 thermal cycles as a function of the ratio $\frac{\tau_{stall}}{\tau_{cycle}}$ (Fig.~\ref{fig:stalling}B, bottom). At $\tau_{stall} = 0$, stalling is not present, but mutant amplification is still suppressed due to collective VCG effects. Mutant amplification drops rapidly as a function of increasing $\tau_{stall}$.

\section{Discussion}

Our results show that collective replication in a virtual circular genome (VCG) architecture intrinsically suppresses the propagation of mutant alleles, even in the absence of high-fidelity copying or kinetic proofreading. Experiments, simulations, and a reduced theory show that suppression is strongest when mutations are 3$^\prime$-proximal and increases with the VCG’s `virtualness' (redundancy of overlapping oligos).  

Intuitively, collective encoding of information builds a directional bias into replication. Alleles that match the consensus retain multiple routes to be replicated, whereas sequence changes (e.g., a mutant allele) prune those routes. We quantified these routes in terms of binding partners since
an allele's rate of being copied depends on how many annealed partners place their 3$^\prime$ end \emph{downstream} of that allele so that extension traverses it. 3$^\prime$-proximal mutants have fewer such productive partners and thus are at a disadvantage compared to wildtype alleles that are coded at varying distances from the 3$^\prime$ end. Stalling at mismatched 3$^\prime$ ends prolongs this period of lowered replicative ability by delaying extension past the mismatch. Consequently, the combined effect of partner limitation and stalling leads to strong suppression that strengthens with virtualness and with increasing ratio $\tau_{stall}/\tau_{cycle}$ of stalling time to melt-anneal-extend cycle time. Unlike mispriming suppression in PCR, which relies on reduced binding affinity from mismatches, our effect persists even if mismatches have negligible impact on binding affinity. 

We focused on the maintenance of information in a structured pool, although prior works often focus on the emergence of catalytic cycles from random sequence pools\cite{toyabe2019cooperative,kudella2021structured}. Prior classic theories of distributed replication (e.g., hypercycles\cite{eigen1971selforganization}) posit cooperative autocatalytic networks that can maintain information, but they have been difficult to realize at the molecular level\cite{ szathmary2013propagation}. In contrast, our work begins from a concrete molecular specification --- a set of overlapping oligos and their allowed anneal–extend interactions --- which makes the system experimentally tractable and tunable. This molecular grounding, rather than an abstract mutually catalyzing graph, allowed us to quantify architecture-driven fidelity. In this aspect, our system shares features with molecularly-defined experimental models\cite{ashkenasy2004design, vaidya2012spontaneous, lincoln2009self,li2013dynamic,jeancolas2021rna, ameta2021darwinian}. 

We expect the core ecological mechanism revealed by our DNA model to generalize beyond the specific conditions studied here, though the quantitative outcomes will depend on the replication chemistry (e.g., in the context of RNA\cite{prywes2016nonenzymatic,ding2023experimental}). Differences in persistence length, strand displacement rates, non-enzymatic extension kinetics will shift quantitative behavior. While we focused on primer extension, templated ligation is natural in distributed architectures \cite{toyabe2019cooperative,kudella2021structured,rosenberger2021self,Burger_2025,matsubara2018kinetic,derr2012prebiotically,tkachenko2015spontaneous,tkachenko2018onset,aoyanagi2025experimental} and enables growth from both 5$^\prime$ and 3$^\prime$ ends. The binding partner effect will have to be studied and quantified in the context of ligation to understand biases intrinsic to that mode of replication. 

In addition to copying errors during primer extension, several other processes can introduce mutations that affect VCG stability. Mismatched annealing\cite{todisco2024rna} can generate transient mispairs that alter extension outcomes, while sequence repeats\cite{chamanian2022computer} may create novel annealing registers, allowing oligomers to extend from unintended regions and producing chimeric products. Chemical lesions such as deamination\cite{duncan1980mutagenic} represent another important source of mutation, and these are not confined to the 3$^\prime$ terminus. Our systematic results on mutation position provide a framework to predict the fate of these diverse error sources within distributed architectures.

Our experiments emphasize primer extension in a fixed pool of oligomers that eventually {become unproductive} by forming long, unmeltable duplexes; sustained replication will require continual regeneration of short oligos (e.g., \textit{de novo} synthesis from activated nucleotides\cite{zhou2021virtual}, {or cleaving oligos into short oligos}\cite{tkachenko2024emergence}). Moving to driven {continuous growth may} strengthen mutant suppression, especially under strong stalling, because wildtype alleles can continue to amplify while 3$^\prime$-proximal mutants are {effectively diluted out before they elongate sufficiently to act as a template}\cite{rajamani2010effect, goppel2021kinetic, matsubara2023avoidance, ichida2005high}. 

Despite these limitations and potential extensions, our work shows that genomic architecture can act as an intrinsic bias for the fate of stored information. That is, by filtering which novelties are propagated versus rejected, the architecture effectively tunes the error threshold without explicit proofreading. The Virtual Circular Genome model was proposed\cite{zhou2021virtual,ding2023experimental} to explain how primitive replication systems could avoid issues associated with the replication of long linear genomes such as strand separation and replication of terminal bases. Surprisingly, we have now shown that it also mitigates the unrelated problem of high intrinsic error rates, and thereby provides a means of avoiding the Eigen error catastrophe\cite{eigen1971selforganization} and thus allowing for the emergence of Darwinian evolution. 

Comparable architecture-level filters might be operating in extant biology, for example how RNA viruses with segmented genomes can undergo genetic reassortment\cite{mcdonald2016reassortment}. 
Other non-standard genome architectures include those of multipartite bacteria (e.g., \textit{Borrelia burgdorferi}\cite{fraser1997genomic,ren2023organization}), which partition essential genes across a chromosome plus several plasmids; ciliates package somatic information into vast numbers of gene-sized macronuclear chromosomes\cite{arnaiz2012paramecium}; 
and trypanosomatids\cite{lukeš2002kinetoplast}, which maintain kinetoplast DNA as a topologically linked network of maxi- and minicircles; and mitochondria and chloroplasts carry tens to thousands of genome copies per cell\cite{chen2005organization}. 
Even among close relatives, karyotypes differ sharply, e.g., \emph{Saccharomyces cerevisiae} has 16 nuclear chromosomes whereas \emph{Schizosaccharomyces pombe} has 3. In all these cases, information is distributed across many molecules with heterogeneous maintenance and inheritance, rather than on one continuous polymer. We lack a general framework for how such architectures influence mutation fate, dosage control, recombination, assortment noise, and long-term evolvability or why particular architectures arise and persist. While we do not address architectural features specific to these extant systems, our work highlights open questions about the advantages of distributing genomic information across many versus few molecules.

\begin{acknowledgments}
\textbf{Acknowledgments.} We thank Irene Chen, Deborah Fygenson, Paul Higgs, Pepijn Moerman, Erik Winfree, and members of the Murugan and Szostak labs for helpful discussions. This work was supported in part by grants from the Sloan Foundation (19518) and the Gordon and Betty Moore Foundation (11479) to J.W.S. and A.M.  J.W.S. is an Investigator of the Howard Hughes Medical Institute. A.M. acknowledges support from NIGMS of the National Institutes of Health under award number R35GM151211. M.J.F. is supported by the Eric and Wendy Schmidt AI in Science Postdoctoral Fellowship, a Schmidt Sciences program. This work was supported by the National Science Foundation through the Center for Living Systems (grant no. 2317138) and by the University of Chicago’s Research Computing Center. 
\end{acknowledgments}

\textbf{Data Availability.} All code to run the simulations is publicly available\cite{falkma_2025_17179043}. Experimental data will be publicly available upon publication.

\bibliographystyle{unsrtnat}
\bibliography{ref}

\setcounter{equation}{0}
\def\theequation{S\arabic{equation}}
\setcounter{figure}{0}
\def\thefigure{S\arabic{figure}}

\end{document}


\def\thefigure{S\arabic{figure}}

\tableofcontents

\section{Sequence Design}\label{sec:seq_design}

We designed a 60 base pair (bp) circular DNA sequence to serve as a virtual circular genome (VCG), with the sequence provided in Supp.~Sec.~\ref{supp:VCG_sequence}. To ensure specific binding of DNA oligomers (oligos) during annealing and prevent unintended interactions, we carefully avoided repeats of subsequences of length four. 

The full VCG comprises 24 oligos -- 12 oligos in one direction and their reverse complements -- mapped onto the 60 bp double-stranded sequence. Each oligo is 25~nt long and overlaps by 20~nt at its 3$^{\prime}$ end with the downstream neighbor. Consequently, each oligo has four possible downstream reverse-complementary neighbors for partial binding.

We analyzed the melting temperatures of all VCG oligos using the NUPACK web server. The original 25~nt VCG oligos exhibited melting temperatures around $61~^\circ\mathrm{C}$, which increased to approximately $80~^\circ\mathrm{C}$ when assuming extension to 40~nt products using other VCG oligos as templates. These metrics indicate that the VCG oligos could separate during thermocycling, in subsequent experiments (Supp.~Sec.~\ref{sec:vcg_cycles}). 

In addition to the core VCG oligos, we synthesized three mutant oligos that differ from one of the VCG oligos (noted as wildtype A1 oligo) by a single, contiguous 4~nt segment (Supp.~Sec.~\ref{sec:SupMat} Table \ref{tab:oligo_sequences}).
When choosing the replacement sequence for each 4~nt block, we required that the new tetranucleotides do not occur anywhere else in either oligo of the 60~bp VCG map that it disrupts local complementarity. Sequence uniqueness ensures that the mutant subsequence cannot form unintended base pairs with any other VCG oligos, thereby preventing spurious oligo interactions during melt-anneal cycles. The three mutant oligos, $\text{A1}^{mut}_{3^\prime-end}$, $\text{A1}^{mut}$, and $\text{A1}^{mut}_{5^\prime-end}$ have the mutant regions at the $3^\prime$ end, the middle, and $5^\prime$ end of the $\text{A1}$ wildtype oligo. The $\text{A1}^{mut}$ with middle mutation was tracked in most experiments, while $\text{A1}^{mut}_{3^\prime-end}$ and $\text{A1}^{mut}_{5^\prime-end}$ were used to investigate the effects of mutation position on mutant replication within the VCG system (Supp.~Sec.~\ref{sec:SupMat} Table \ref{tab:oligo_sequences}).

All oligos, desalting purified, were purchased as dry from Integrated DNA Technologies (Coralville, IA), and resuspended using Milli-Q water to 100~uM.

\section{VCG-Mutant Mixture Compositions }
\label{VCG mix}

We prepared mixtures of different VCG and mutant components to study three important features of VCG-mutant system that potentially influence the competitive amplification dynamics between wildtype and mutant sequences: 

\begin{enumerate}
\item \textbf{VCG Virtualness}: We created three versions of the VCG construct, {VCG12}, VCG6, and VCG3 with 12, 6, and 3 oligos plus their reverse complementary oligos. While all mixtures cover the entire VCG, they do so with different levels of coverage and overlap. See Supp.~Sec.~\ref{sec:SupMat} Table \ref{tab:def_virtualness} for oligo components of these mixtures. 

\item \textbf{Mutation Position on the A1 oligo}: We introduced 3 mutants of {A1} wildtype oligo ($\text{A1}^{mut}_{3^\prime-end}$, $\text{A1}^{mut}$, and $\text{A1}^{mut}_{5^\prime-end}$, sequences included in Supp.~Sec.~\ref{sec:SupMat} Table \ref{tab:oligo_sequences}) whose mutations locate near the $5^\prime$ end, middle, or $3^\prime$ end of the A1 oligo to evaluate the impact of mutation position on mutant propagation and competitive suppression within the VCG system. 

\item \textbf{Starting Mutant Proportions}: The proportion of mutants is defined as the ratio of mutant concentration introduced to the A1 wildtype oligo concentration. Three different proportions -- 2.5\%, 5\%, or 25\% -- were used to study how initial mutant proportions in VCG-mutant mixtures affect mutant proliferation dynamics. 
\end{enumerate}

In the experiments described below, we mixed VCGs of different virtualness (VCG12, VCG6, or VCG3) with varied proportions (2.5\%, 5\%, or 25\%) of mutants with varied mutation positions ($\text{A1}^{mut}_{3^\prime-end}$, $\text{A1}^{mut}$, or $\text{A1}^{mut}_{5^\prime-end}$). 

\section{VCG Cycles of Melt, Anneal, and Extension by  Thermocycling}\label{sec:vcg_cycles}

\subsection{Melt-anneal-extend cycles of the VCG with Bst enzyme}

Primer extension reactions were performed using {Bst DNA Polymerase, Large Fragment} (New England Biolabs, Cat.~No.~\textit{M0275}) on the {Eppendorf\texttrademark~Mastercycler\texttrademark~pro PCR System}.

\begin{table}[h!]
\centering
\begin{tabular}{@{}lll@{}}
\toprule
\textbf{Component} & \textbf{Volume (uL)} & \textbf{Final Concentration} \\
\midrule
VCG Oligo Mix& 2.00  & 2~uM \\
10$\times$ Buffer                         & 0.50  & 1$\times$ \\
Mg\textsuperscript{2+} (100~mM stock)       & 0.25  & 5~mM \\
dNTPs (10~mM stock)                & 0.50  & 1~mM \\
A1\textsuperscript{mut}           & 1.00  & 2.5\%, 5\%, or 25\% of VCG conc. \\
Bst DNA Polymerase (8000~U/mL)    & 0.20  & 320~U/mL \\
Nuclease-free H\textsubscript{2}O        & 0.55  & --- \\
\midrule
\textbf{Total}                    & \textbf{5.00} & --- \\
\bottomrule
\end{tabular}
\end{table}

Here, as described in Supp.~Sec.~\ref{VCG mix}, VCG oligos were mixed at a 1:1 stoichiometry, consisting of 12, 6, or 3 oligos plus their reverse complementaries for VCG12, VCG6, or VCG3. Mutant oligos ($\text{A1}^{mut}_{3^\prime-end}$, $\text{A1}^{mut}$, or $\text{A1}^{mut}_{5^\prime-end}$) are introduced at various initial proportions ranging from 2.5\% to 25\% (Supp.~Sec.~\ref{VCG mix}) of the VCG mix concentration. In the table above, the 2~uM final VCG concentration refers to each oligo's concentration composing the VCG mix. 

We used the following thermocycling protocol. The number of Denaturation and Annealing/Extension cycles was varied between 10, 20, or 30 as indicated. We also carried out a `0 cycle' control in which the mixture was prepared exactly as in the table above for the 10, 20, and 30 cycle experiments; the mixture was subject to only the Enzyme Deactivation step ($90~^\circ\mathrm{C}$, 10~min) in the table below. 

\begin{table}[h!]
\centering
\begin{tabular}{@{}llll@{}}
\toprule
\textbf{Step}              & \textbf{Temperature} & \textbf{Duration} & \textbf{Cycles} \\
\midrule
Denaturation               & $80~^\circ\mathrm{C}$     & 30~s        & 10, 20, or 30 \\
Annealing/Extension        & $35~^\circ\mathrm{C}$     & 1~min          & 10, 20, or 30 \\
Final Extension            & $35~^\circ\mathrm{C}$     & 2~min         & 1 \\
Enzyme Deactivation        & $90~^\circ\mathrm{C}$     & 10~min        & 1 \\
Hold                       & $4~^\circ\mathrm{C}$                 & until use         & --- \\
\bottomrule
\end{tabular}
\end{table}

\section{qPCR quantification of wildtype and mutant abundance} \label{sec:qpcr_description}

To quantify the relative amounts of A1 and $\text{A1}^{mut}$ in oligo mixtures, we employed quantitative PCR (qPCR) assays using two sets of primers specifically designed to bind either A1 or $\text{A1}^{mut}$ (Sequences in Supp.~Sec.~\ref{sec:SupMat} Table \ref{tab:primer_sequences}). These primers allowed us to resolve the growth of each allele sequence following VCG extension. 

The design enables discrimination between A1 and $\text{A1}^{mut}$ by using a common reverse primer (Common\_rev) that anneals to the reverse-complementary sequences of both A1 and $\text{A1}^{mut}$, and distinct forward primers ($\text{A1}_{fw}$ and $\text{A1}^{mut}_{fw}$) whose 3$^\prime$ ends anneal to the sequence regions where $\text{A1}^{mut}$ differs from A1. Similar primer design logic was applied to $\text{A1}^{mut}_{3^\prime-end}$ and $\text{A1}^{mut}_{5^\prime-end}$. All primer sequences are provided in Supp.~Sec.~\ref{supp:primer_sequences}.

qPCR assays were performed using the QuantStudio 7 Pro Real-Time PCR System (Applied Biosystems) and Power SYBR Green Master Mix (ThermoFisher, Cat.~No.~\textit{4367659}). Each qPCR reaction has a total volume of 4~uL arranged in 384-well plates. qPCR assay compositions and thermocycling programs are summarized in the following tables. Each sample was run in duplicate.

\begin{table}[h!]
\centering
\begin{tabular}{@{}lll@{}}
\toprule
\textbf{Component}                               & \textbf{Volume (uL)} & \textbf{Final Concentration}           \\
\midrule
Power SYBR Green PCR Master Mix (2$\times$)            & 2.00                 & 1$\times$                                   \\
Forward primer (specific)                        & 0.04                & 500~nM                                \\
Reverse primer (specific)                        & 0.04                & 500~nM                                \\
Template (A1, $\text{A1}^{mut}$, or VCG)   & variable\textsuperscript{a} & 0.02--1~nM, see Sec.~\ref{sec:qpcr_CrossValidation}          \\
Nuclease-free H$_2$O                             & to 4.00             & ---                                     \\
\midrule
\textbf{Total}                                   & \textbf{4.00}        & ----                                     \\
\bottomrule
\end{tabular}

\smallskip
\begin{minipage}{0.9\linewidth}\footnotesize
\textsuperscript{a}Template input was serially diluted to span the concentration range established in Supp.~Sec.~\ref{sec:qpcr_CrossValidation}.
\end{minipage}
\end{table}

\begin{table}[h!]
\centering
\begin{tabular}{@{}llll@{}}
\toprule
\textbf{Step}            & \textbf{Temperature} & \textbf{Duration} & \textbf{Cycles} \\
\midrule
Initial denaturation     & $95~^\circ\mathrm{C}$               & 10~min            & 1              \\
Denaturation             & $95~^\circ\mathrm{C}$               & 15~s              & 40             \\
Annealing/extension      & $55~^\circ\mathrm{C}$               & 60~s              & 40             \\
\bottomrule
\end{tabular}
\end{table}

Each qPCR assay aimed to detect the relative abundance of A1 and $\text{A1}^{mut}$ in a sample. Samples were divided into two aliquots: one received the A1-specific primer set, and the other received the $\text{A1}^{mut}$-specific primer set. Delta normalized reporter signal ($\Delta \text{Rn}$) was collected throughout the thermocycling process, and threshold cycle (Ct) values were determined using a constant $\Delta \text{Rn}$ threshold.

\subsection{Cross-Validation of Primer Specificity and Sensitivity in qPCR}
\label{sec:qpcr_CrossValidation}

To ensure accurate quantification of the relative abundances of wildtype (A1) and mutant templates ($\text{A1}^{mut}_{3^\prime-end}$, $\text{A1}^{mut}$, or $\text{A1}^{mut}_{5^\prime-end}$) in our VCG system, we performed cross-validation assays assessing primer-template specificity and sensitivity. The accuracy and robustness of qPCR quantitation critically depend on these parameters; thus, the validation procedure was structured to evaluate primer-template interactions under both matched and mismatched conditions.

\subsubsection{Matched Primer-Template Specificity}
\begin{figure}[H]
\includegraphics[width=\linewidth]{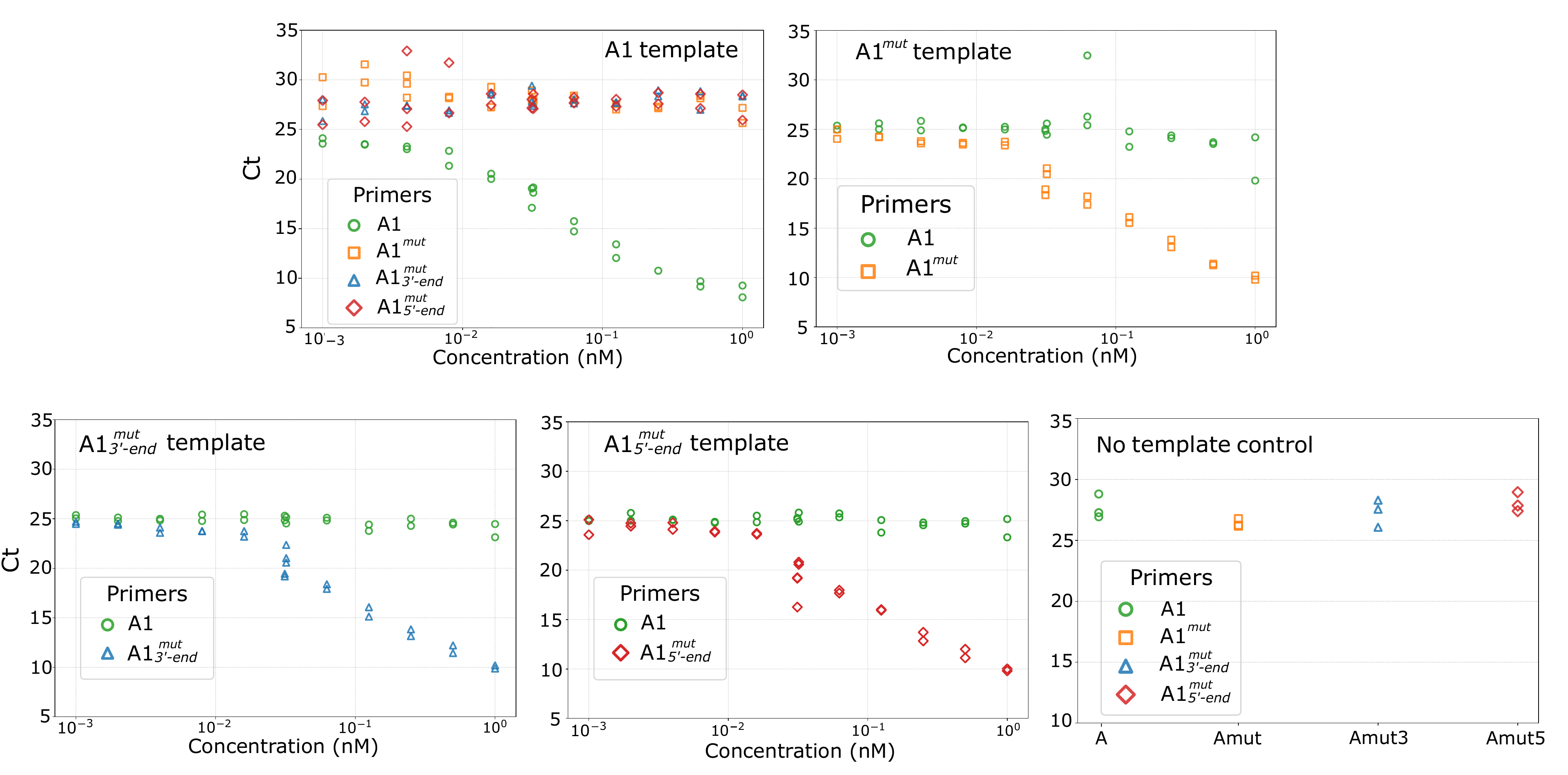}
\caption{\textbf{Evaluation of primer-template specificity under mismatched conditions.} Primer concentrations were maintained at 500~nM. The substantially delayed amplification (high Ct) confirms minimal nonspecific interactions between mismatched primers and templates.}
\label{fig:mismatching-CV}
\end{figure}

We first established the quantitative sensitivity window for each perfectly matched primer-template pair (Fig.~\ref{fig:mismatching-CV}).  In the range from roughly 0.01 to 1 nM template, two-fold serial dilutions produced clear, concentration-dependent shifts in cycle threshold (Ct) values for all four targets (A1, $\text{A1}^{mut}_{3^\prime-end}$, $\text{A1}^{mut}$, or $\text{A1}^{mut}_{5^\prime-end}$).  These well-defined Ct gradients demonstrate efficient primer hybridization and exponential amplification, confirming the robust sensitivity of the qPCR system to accurately quantify each VCG oligo.  

We further evaluated mismatched primer-template combinations to assess nonspecific amplification (Fig.~\ref{fig:mismatching-CV}). For each template (A1, $\text{A1}^{mut}_{3^\prime-end}$, $\text{A1}^{mut}$, or $\text{A1}^{mut}_{5^\prime-end}$), only the corresponding matched primer generated efficient amplification, and the calculated Ct values depend on the template concentrations tested. In contrast, all mismatched primers produced high Ct values ($\ge$25) across the entire range of template concentrations tested (down to 1~pM), reflecting poor hybridization and negligible amplification. These results were consistent with the no-template control (NTC), which also yielded Ct values in the 25--30 range, possibly attributable to low-level primer-dimer formation.

Critically, we observed that even matched primer-template pairs become indistinguishable from background when template concentrations fall below 0.01~nM. In this range, Ct values plateau around 25--30, converging with those of mismatched pairs and the NTC. This convergence establishes a practical lower detection limit ($\sim$10~pM) for reliable qPCR resolution of target template concentration.

Together, these cross-validation data confirm that our qPCR system achieves high specificity and sensitivity across a defined operational window. Above the lower detection limit threshold, it robustly distinguishes wildtype and mutant allele sequences, even within complex mixtures of closely related oligos.

\subsubsection{Mismatched Primer-Template Specificity}

\section{Echo Robot workflow for Bst and qPCR setup}

For systematic investigation of VCG extension under diverse conditions, we utilized an Echo 525 acoustic liquid handler (Beckman Coulter) for precise and efficient setup of numerous small-volume reactions. VCG oligos and Bst reagents are placed in separate wells of an Echo source plate. For each thermal cycle (0, 10, 20, or 30), a 96-well destination plate was prepared, each well containing a 5~µL Bst polymerase reaction mixture. Each reaction was formulated to vary systematically in VCG mix composition, allowing explorations of different VCG system properties. 

To perform qPCR assay on the 96-well Bst reaction plate as described in Supp.~Sec.~\ref{sec:vcg_cycles},  Bst reaction products from the 96-well plate are first diluted by MINI 96 96-channel Portable Electronic Pipette (INTEGRA Biosciences) and transferred to an Echo 384-well source plate. This step dilutes and aliquots each Bst reaction from the 96-well plate into four wells of the 384-well plate, ready for qPCR assays. Dilution levels were determined from pilot control tests (Supp.~Sec.~\ref{sec:qpcr_CrossValidation}). With the Power SYBR Green master mix and primers that are prepared separately in an Echo source plate, we use the Echo to set up duplicated, parallel qPCR assays of A1 and $\text{A1}^{mut}$ as described in Supp.~Sec.~\ref{sec:qpcr_description}.

\section{Main Experimental Results}\label{sec:mutant_suppression}

\subsection{Experimental setup}

As described above (Supp.~Sec.~\ref{sec:vcg_cycles} and~\ref{sec:qpcr_description}), our experiments comprised three principal stages:

\begin{enumerate}
\item \textbf{VCG Extension via Bst Thermocycling:}
Reaction mixtures were prepared with defined oligo compositions, specifically varying in two parameters: (a) the virtualness of the VCG (VCG12, VCG6, or VCG3. Sequences in Supp.~Sec.~\ref{sec:SupMat} Table \ref{tab:def_virtualness}) and (b) the position of introduced mutations on the $\text{A1}^{mut}$ oligo ($\text{A1}^{mut}_{3^\prime-end}$, $\text{A1}^{mut}$, or $\text{A1}^{mut}_{5^\prime-end}$ as in Supp.~Sec.~\ref{sec:SupMat} Table 
\ref{tab:oligo_sequences}). 
Extension of the VCG-mutant mixtures was carried out by the thermal cycle in duplicates as described in Supp.~Sec.~\ref{sec:vcg_cycles}.  
\item \textbf{Sample Dilution and Preparation for qPCR:}  
Immediately following Bst-mediated thermocycling, extended VCG-mutant sample was diluted in nuclease-free water to ensure that template concentrations fell within the empirically determined dynamic range ($\approx$ 0.01--1~nM) established in cross-validation assays for qPCR (Supp.~Sec.~\ref{sec:qpcr_CrossValidation}). SYBR Green-based qPCR was then arranged as described in Supp.~Sec.~\ref{sec:qpcr_description}, with each Bst product split to two groups of aliquots, in duplicates, designated for primers specific to the A1 wildtype and mutants. 
\item \textbf{Quantitative qPCR Analysis:}  
From qPCR runs, threshold cycle (Ct) values were recorded and analyzed to calculate amplification rates of A1 and $\text{A1}^{mut}$, details included in the following
 Supp.~Sec.~\ref{sec:mutant_suppression}B. 
\end{enumerate}

\subsection{Data Analysis}

\begin{figure}
\includegraphics[width=\linewidth]{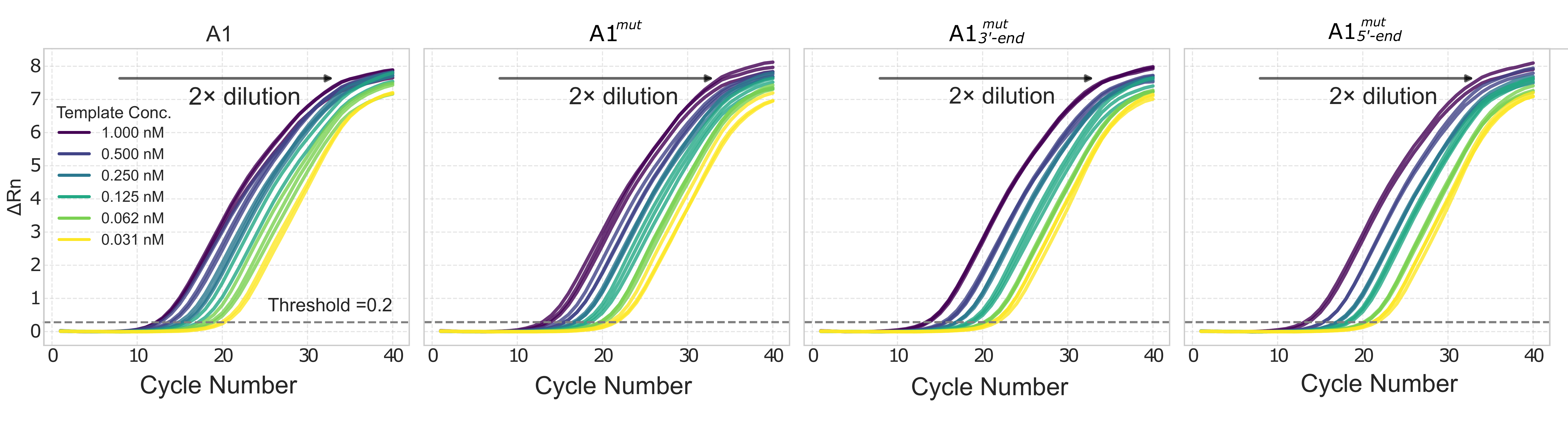}
\includegraphics[width=\linewidth]{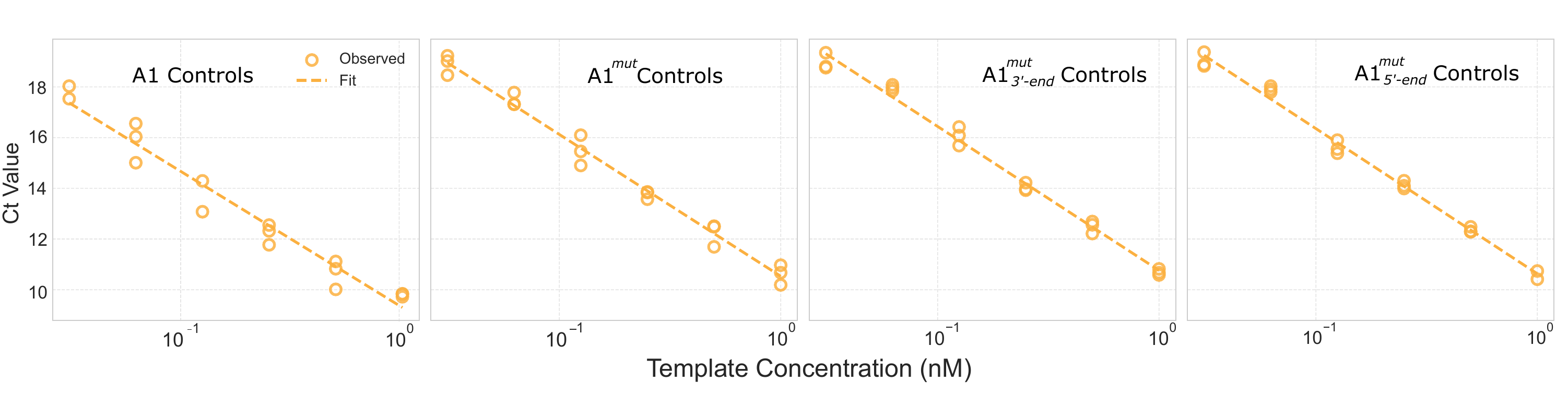}
\caption{\textbf{Calibration of Ct versus template concentration for Ct-concentration conversion.} 
Top panel: Example $\Delta \text{Rn}$ amplification curves for matched primer-template pairs (A1, $\text{A1}^{mut}_{3^\prime-end}$, $\text{A1}^{mut}$, and $\text{A1}^{mut}_{5^\prime-end}$), showing clear, concentration-dependent rightward shifts in cycle threshold. Templates were serially diluted two-fold from 0.5~nM (A1) or 1~nM (all mutants). The horizontal dashed line at $\Delta \text{Rn}$ = 0.2 marks the analysis threshold. 
Bottom panel: Standard curves plotting measured Ct against $\log_{10}(\text{template concentration})$. Linear regression yields slope (m) and intercept (b) values used to convert experimental Ct measurements into absolute template concentrations in subsequent assays.}
\label{fig:FinalExp-Control}
\end{figure}

We calibrated the Ct values of each sample by fitting the observed Ct values to a calibration curve established by control A1 or $\text{A1}^{mut}$ templates of serially diluted concentrations, as in Fig.~\ref{fig:FinalExp-Control}. The concentration range of the controls is informed by the sensitive qPCR detection range for A1 and $\text{A1}^{mut}$ signals as tested in Supp.~Sec.~\ref{sec:qpcr_CrossValidation} 

The calibration curve (Fig.~\ref{fig:FinalExp-Control}) supports that the observed Ct values are inversely proportional to the logarithm of the actual concentration of the target sequence, expressed as:
\[
\text{Ct} = m \times \log_{10}(\text{concentration}) + b,
\]
where $m$ is a constant and $b$ is the intercept. 
The amplification status of A1 and $\text{A1}^{mut}$ during VCG extension was determined by comparing their relative concentrations at different cycles. The fold-change in concentration was calculated as:
\[
[\text{A1 or }\text{A1}^{mut}]_t = 10^{\frac{\text{Ct}_t - b}{m}},
\]
where $\text{Ct}_t$ are the Ct values at cycles $t$ = $\{0, 10, 20,30\}$. This is used to represent the normalized amplifications of A1 and $\text{A1}^{mut}$ at each thermal cycle. 

The wildtype advantage compared with the mutant is further characterized as the ratio of concentration between the wildtype and mutant at cycle 30:
\[
w = \frac{[\text{A1}]_{30}}{[\text{A1}^{mut}]_{30}}.
\]

\begin{figure}[H]
  \centering
  \includegraphics[width=\linewidth]{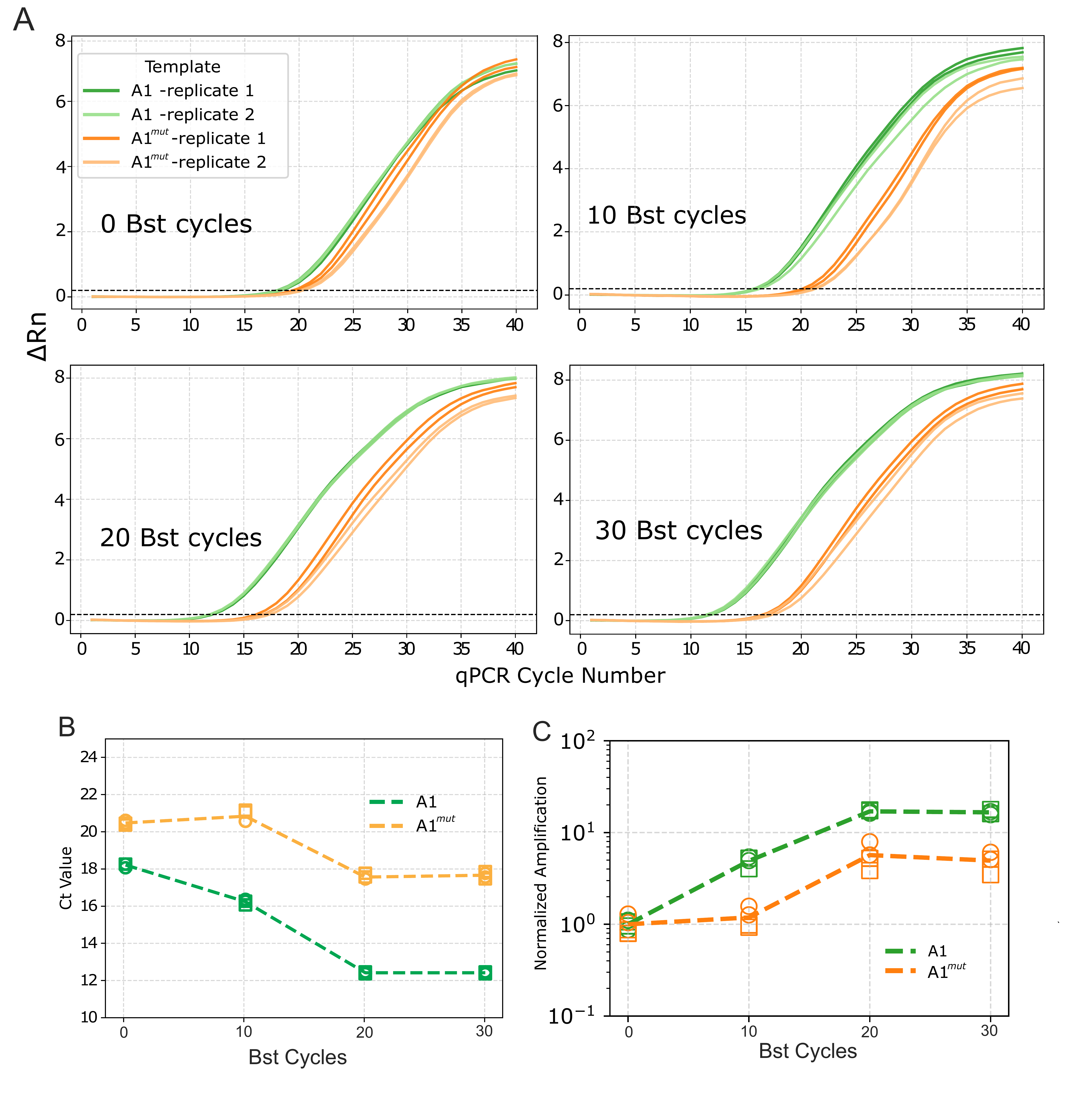}
  \caption{%
    \textbf{Example amplification curves and normalized amplification for 5\% $\text{A1}^{mut}$ (middle mutation) in VCG12.}  
    (A) Raw $\Delta \text{Rn}$ qPCR traces for the A1 and $\text{A1}^{mut}$ signals after 0, 10, 20, or 30 Bst extension cycles. The dashed line at $\Delta \text{Rn}=0.2$ denotes the analysis threshold.  
    (B) Extracted Ct values for A1 and $\text{A1}^{mut}$ based on the amplification curves in (A).  
    (C) Normalized amplification of A1 and $\text{A1}^{mut}$ -- calculated as described in Supp.~Sec.~\ref{sec:mutant_suppression}B. Symbols show independent replicates; dashed lines indicate their mean.%
}
  \label{fig:ExampleAmp}
\end{figure}

Fig.~\ref{fig:ExampleAmp} shows an example of our analysis pipeline for the VCG12 reaction containing 5\% $\text{A1}^{mut}$ (middle-position mutation). Extended VCG mixtures were taken after 0, 10, 20, or 30 cycles of Bst extension and separately subjected to SYBR green qPCR as described above in Supp.~Sec.~\ref{sec:vcg_cycles} and \ref{sec:qpcr_description}. For each thermal cycle, we extracted the cycle threshold (Ct) from the raw $\Delta\text{Rn}$ curves using a threshold of 0.2. These Ct values were then converted into fold-change of template concentrations for both A1 and $\text{A1}^{mut}$ by reference to calibration controls for each qPCR run (e.g., the qPCR controls for 10-cycle Bst samples in Fig.~\ref{fig:FinalExp-Control}). Applying this procedure across all four thermal cycle points produces the amplification trajectories plotted in Fig.~\ref{fig:ExampleAmp}, which illustrate how wildtype and mutant species accumulate over 0--30 cycles under the VCG12, 5\% $\text{A1}^{mut}$ (middle-position) condition. The advantage of wildtype A1 over the mutant $\text{A1}^{mut}$, denoted by $w$, is further summarized in Fig.~\ref{fig:RelativeFitness}.

\subsection{Result: Effect of VCG Virtualness on Mutant Suppression}

We prepared reaction setups with three distinct virtualness conditions (VCG12, VCG6, and VCG3 as described in Supp.~Sec.~\ref{sec:mutant_suppression}A).

Fig.~S4--S6 presents the full experiment analysis for 2.5\%, 5\%, and 25\% initial mutants. Summaries of the wildtype advantage regulated by different VCG-mutant conditions are plotted in Fig.~\ref{fig:RelativeFitness}. Under each condition tested, there is a consistent trend of growth for both A1 and $\text{A1}^{mut}$ across the 30 thermal cycles at different rates. The growth is evident in the first 20 thermal cycles and reach a plateau after around 20 cycles, consistent with the interpretation of VCG extension saturation from the gel image (main text Fig.~2D).

Across all initial mutant fractions (2.5\%, 5\%, or 25\%) and for each mutation position, higher VCG virtualness produces dramatically stronger suppression of mutant amplification relative to the wildtype A1 signal. 
For instance, consider the 2.5\% $\text{A1}^{mut}_{3^\prime-end}$ data in Fig.~\ref{fig:FinalAmpPanel2.5}.  
In the high-virtualness VCG12 mixture (top row, right-most column), the normalized A1 (wt) amplification rises to approximately 7$\times$ at 10 cycles, and then to about 15$\times$ at 20 cycles -- where it plateaus -- whereas $\text{A1}^{mut}_{3^\prime-end}$ remains essentially at 1× throughout.  
Hence by cycle 30 the wildtype advantage is roughly 15 (Fig.~\ref{fig:RelativeFitness}). 
By contrast, in the low-virtualness VCG3 mixture (bottom row of Fig.~\ref{fig:FinalAmpPanel2.5}), both A1 and $\text{A1}^{mut}_{3^\prime-end}$ experience notable growths. 
At cycle 30, both A1 and $\text{A1}^{mut}_{3^\prime-end}$ reach about 10$\times$ of initial abundance, leading to a much lower wildtype advantage close to 1. 
Similarly, these trends persist at higher mutant inputs (see Fig.~\ref{fig:FinalAmpPanel5} and \ref{fig:FinalAmpPanel25}).
Taken together, these data quantitatively demonstrate that increasing VCG virtualness -- from VCG3 to VCG12 -- suppresses mutant amplification.

\subsection{Result: Effect of Mutation Position on Mutant Suppression}

Using the three positional mutants $\text{A1}^{mut}_{3^\prime-end}$, $\text{A1}^{mut}$, and $\text{A1}^{mut}_{5^\prime-end}$ as characterized in Sec.~\ref{sec:mutant_suppression} A, we further explored how the positional context of mutations influences its suppression relative to the wildtype A1 sequence. 

\begin{figure}[H]
\centering
\includegraphics[width=\linewidth]{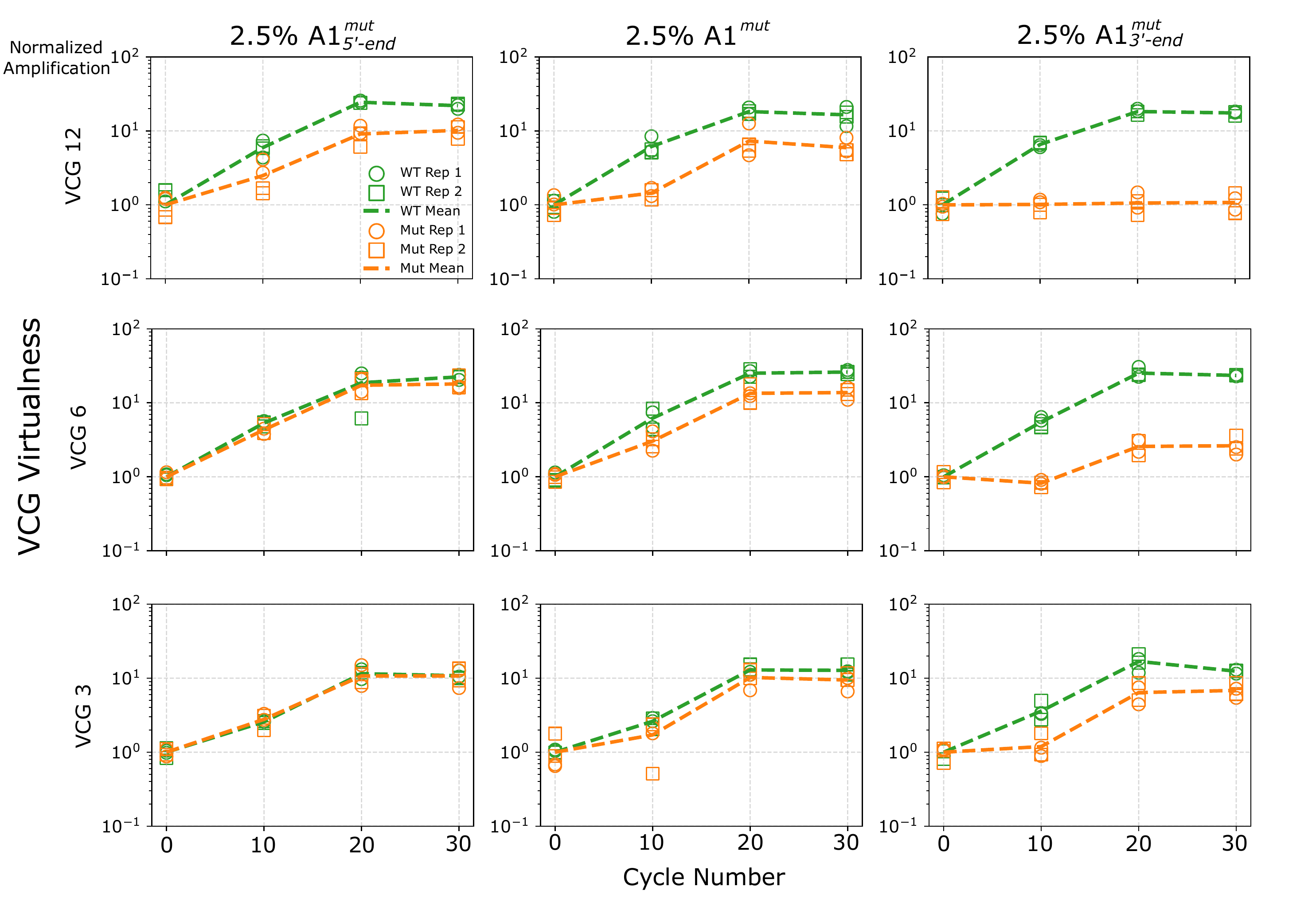}
\caption{\textbf{Normalized amplification across different VCG virtualness and mutation positions under 2.5\% mutant.} Each subplot shows amplification over thermal cycles (0, 10, 20, 30) comparing wildtype (wt, green) and 2.5\% mutant (mut, orange) across VCG virtualness (rows: VCG12, VCG6, VCG3) and mutation positions (columns: $5^\prime$, middle, $3^\prime$). Individual replicates are marked (circles/squares); means are dashed.}
\label{fig:FinalAmpPanel2.5}
\end{figure}

\begin{figure}[H]
\centering
\includegraphics[width=\linewidth]{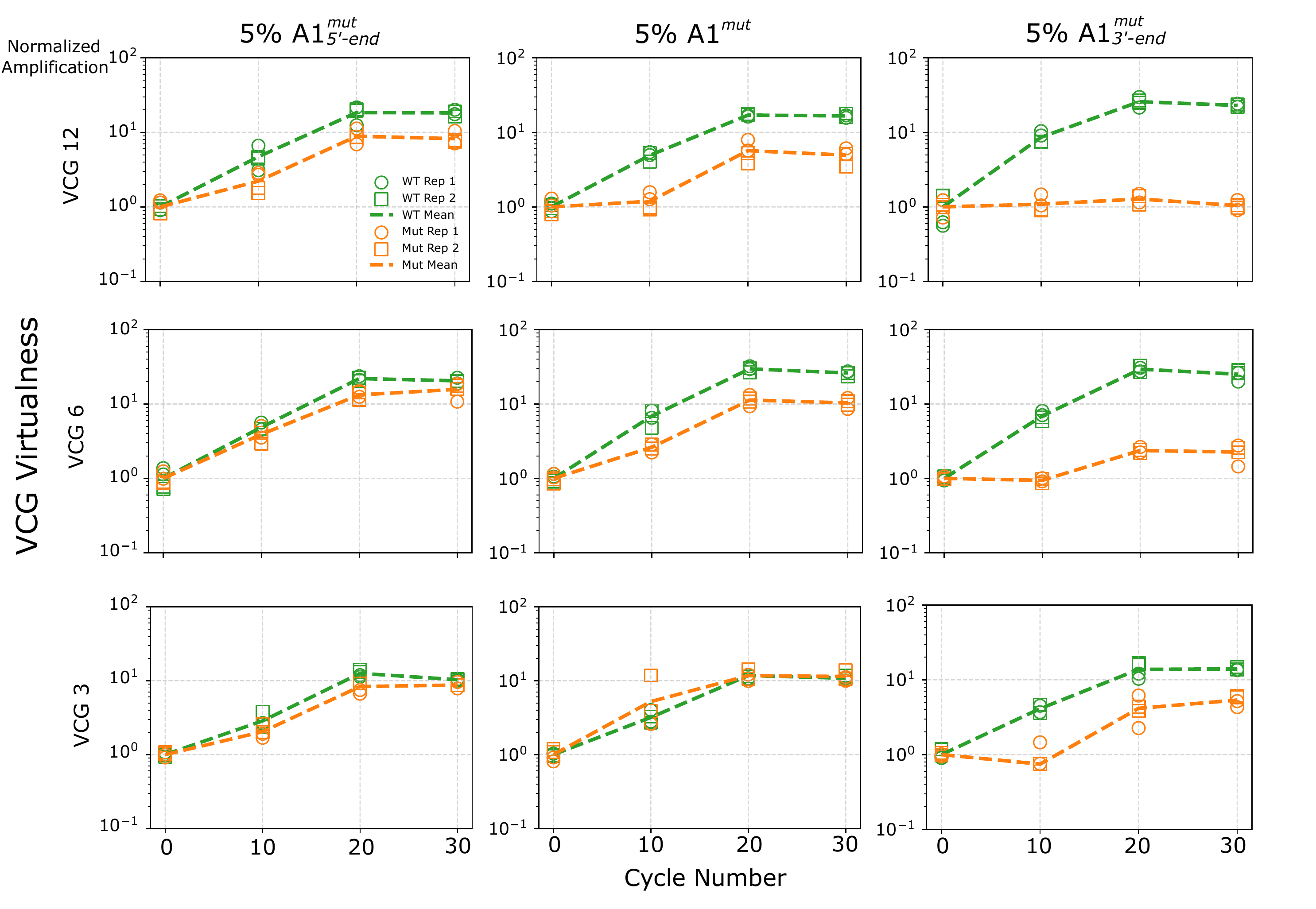}
\caption{\textbf{Similar panel structure and annotations as in Fig.~\ref{fig:FinalAmpPanel2.5}. Normalized amplification across different VCG virtualness and mutation positions under 5\% mutant.}  Each subplot shows amplification over thermal cycles (0, 10, 20, 30) comparing wildtype (wt, green) and 2.5\% mutant (mut, orange) across VCG complexities (rows: VCG12, VCG6, VCG3) and mutation positions (columns: $5^\prime$, middle, $3^\prime$). Individual replicates are marked (circles/squares); means are dashed.}
\label{fig:FinalAmpPanel5}
\end{figure}

\begin{figure}[H]
\centering
\includegraphics[width=\linewidth]{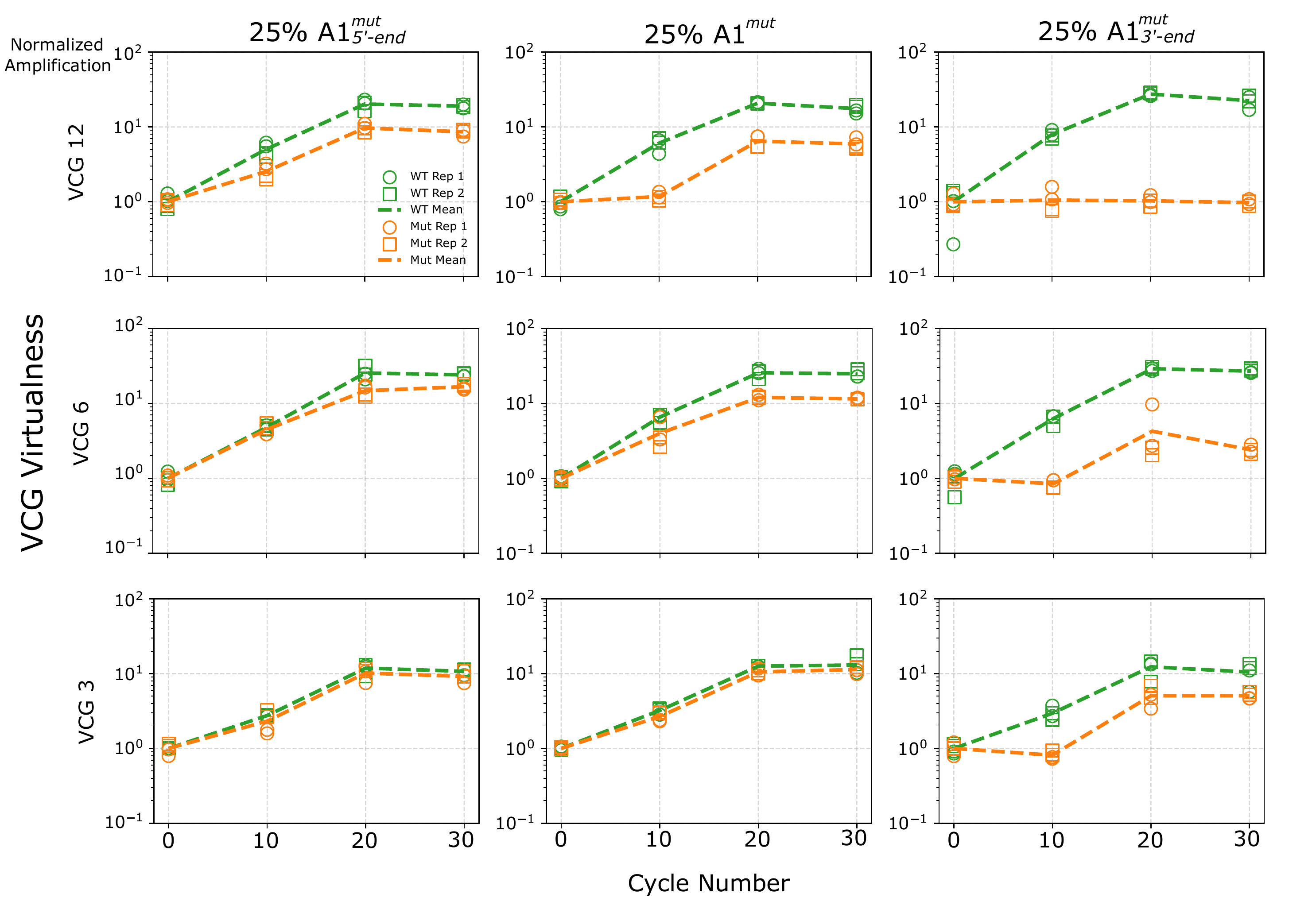}
\caption{\textbf{Similar panel structure and annotations as in Fig.~\ref{fig:FinalAmpPanel2.5}. Normalized amplification across different VCG virtualness and mutation positions under 25\% mutant.}  Each subplot shows amplification over thermal cycles (0, 10, 20, 30) comparing wildtype (wt, green) and 2.5\% mutant (mut, orange) across VCG complexities (rows: VCG12, VCG6, VCG3) and mutation positions (columns: $5^\prime$, middle, $3^\prime$). Individual replicates are marked (circles/squares); means are dashed.}
\label{fig:FinalAmpPanel25}
\end{figure}

\begin{figure} [H]
\includegraphics[width=\linewidth]{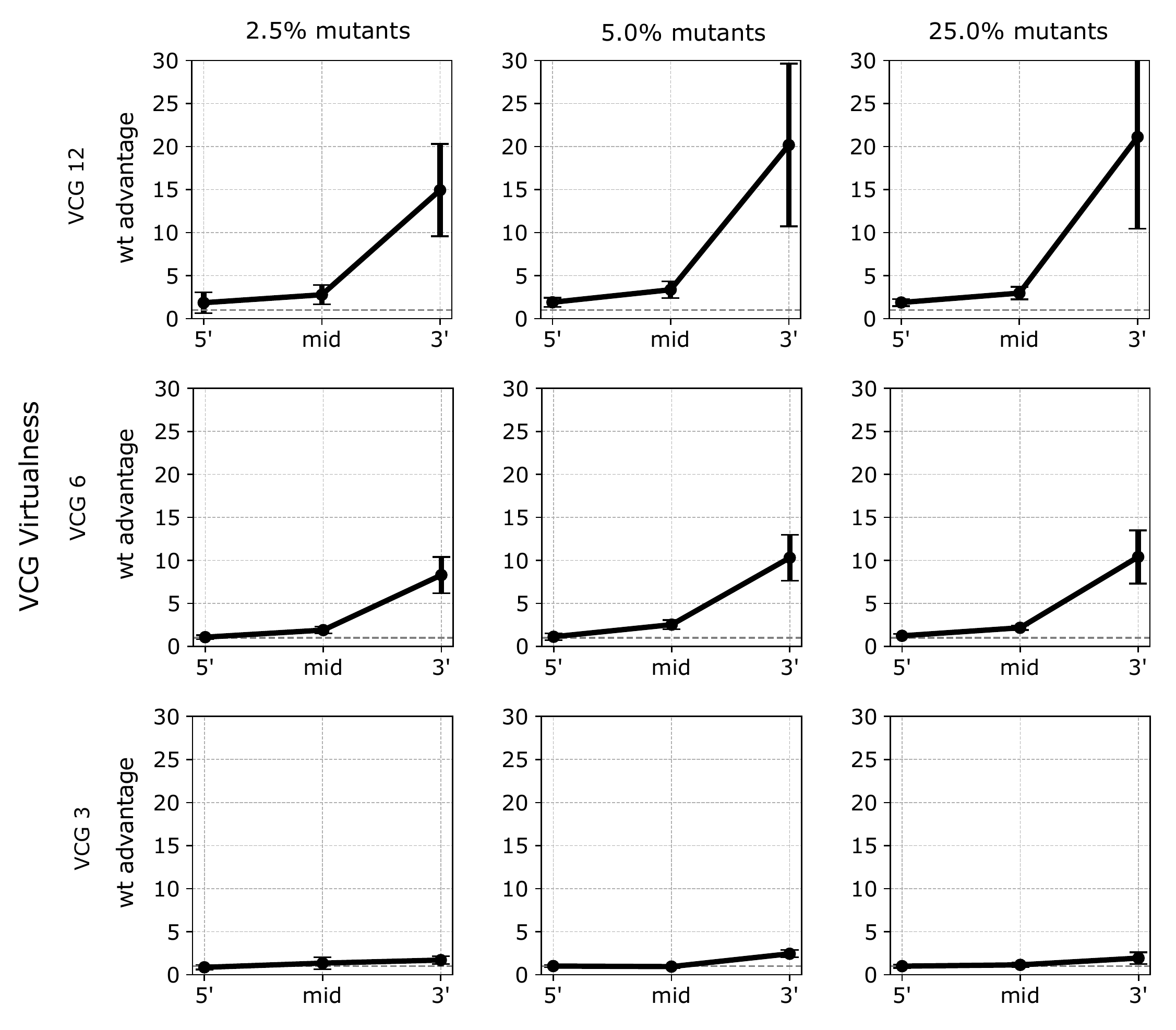}
\caption{\textbf{Wildtype advantage across mutation positions, virtualness levels, and initial mutant fractions.} Each subplot quantifies the amplification advantage of wildtype (A1) over mutant allele sequences ($\text{A1}^{mut}$) by computing the ratio of final amplification levels after 30 thermal cycles. Columns correspond to different initial mutant concentrations (2.5\%, 5\%, and 25\%), and rows represent VCG virtualness levels (VCG12, VCG6, and VCG3). Within each panel, the wildtype advantage is plotted against the mutation position (5$^\prime$, middle, and 3$^\prime$) on the mutant oligo. A consistent trend is observed: the wildtype advantage increases as the mutation shifts toward the 3$^\prime$ end, particularly under high virtualness conditions. This indicates that mutations proximal to the 3$^\prime$ end are more strongly suppressed, likely due to interference with critical neighbor-oligo binding during extension. The effect is robust across varying initial mutant frequencies, suggesting that virtual genome topology -- rather than mutant abundance -- dominantly governs suppression dynamics.}
\label{fig:RelativeFitness}
\end{figure}

From Fig.~\ref{fig:FinalAmpPanel2.5}--\ref{fig:FinalAmpPanel25}, across all conditions, the $5^\prime$ end mutant $\text{A1}^{mut}_{5^\prime-end}$ exhibits the weakest suppression. For instance, at 2.5\% initial proportion,  its amplification curves closely parallel the wt signal under VCG6 and VCG3, yielding the wildtype advantage near unity. At VCG12, $\text{A1}^{mut}_{5^\prime-end}$ mutant signal grows only slightly slower than A1, with the wildtype advantage slightly above 1 (Fig.~\ref{fig:RelativeFitness}).  By contrast, the $3^\prime$ end mutant $\text{A1}^{mut}_{3^\prime-end}$ is much more strongly suppressed: particularly at VCG6 and VCG12,  its normalized amplification remains near 1--2$\times$ while wt rises $\sim\!10$--$15×$, corresponding to the wildtype advantage of around 7 and 15 (Fig.~\ref{fig:RelativeFitness}, left column). 

Together, our data demonstrate that mutation-position influences suppression: mutations near the $3^\prime$ end incur the greatest amplification disadvantage.

\subsection{Result: Effect of Initial Mutant Proportion on Mutant Suppression}

Among Fig.~S4--S6, similar patterns of A1 and $\text{A1}^{mut}$ amplifications regulated by VCG virtualness and mutant position can be observed under different initial mutant proportions from 2.5\% to 25\%. All the wildtype advantage values, regulated by VCG virtualness and mutant position, reported in Fig.~\ref{fig:RelativeFitness}  also agrees quantitatively across all three initial mutant proportion level. This reveals that, within this range, initial proportions of mutant allele sequences in the VCG pool do not notably alter the extent of mutant suppression.  

\section{Simulation Details}

In order to gain further understanding of the mutant suppression effect we observe in our DNA virtual circular genome, we built a simplified simulation model of melt, anneal, and extend cycles which drive extension in a pool of oligos.

\subsection{Sequence representation}

In our simulation, we represent the sequence of the virtual circular genome (VCG) as a segment of integers $[0, L]$. Oligos are represented as directed contiguous subsegments of the VCG segment, potentially up to and including all possible segments of length $L$. The direction of an oligo segment can either be clockwise or counterclockwise, with periodic boundary conditions allowing for oligos which wrap around the VCG. In addition, there is a special position $x_{mut}$ along the genome (fixed here to be $x_{mut} = \lfloor(L/2)\rfloor$) at which an oligo can either have a consensus wt base pair, or contain a mutant base pair. If the oligo has the mutant base pair at $x_{mut}$, then it is considered to be a mutant oligo. 

Therefore, in our simulation, an oligo $i$ can be represented as a 4-tuple with the following data: $(s_i, e_i, h_i,m_i)$, where $s_i$ is the start of the oligo, $e_i$ is the end of the oligo, $h_i$ is the handedness of the oligo (Boolean 0 or 1 for clockwise or counterclockwise), and $m_i$ is the mutant status of the oligo (Boolean 0 or 1 for mutant or non-mutant). Each of these 4-tuples is indexed by a unique integer ($i$) and carries with it a concentration $c_i$. 

Our simulation also needs to record duplexes that form when two oligos anneal to each other. Duplexes can be indicated as pairs of indices $(i, j)$ (shorthand $ij$) corresponding to two single-stranded oligos of opposing direction. To distinguish between the ambiguity of a duplex $ij$ versus $ji$, the convention is that the clockwise oligo index is listed first. Each duplex has an associated concentration $c_{ij}$.

\subsection{Annealing}

During annealing, single-stranded oligos in the VCG pool react to become duplexes via irreversible second-order kinetics $i ~+~ j \rightarrow ij $ between oligos $i$ and $j$ to create duplex $ij$. The corresponding ODE is given as:
\begin{equation}
    \frac{dc_{ij}}{dt} = k_{anneal} \Theta(o_{ij}-o_{min})(1-\delta_{h_ih_j})c_ic_j, 
\end{equation}
where $\Theta$ indicates a Heaviside step function, $\delta$ indicates a Kronecker delta, $k_{anneal}$ is an overall reaction rate, $o_{ij}$ is the overlap between the segments of oligos $i$ and $j$, $o_{min}$ is a minimal overlap parameter necessary for annealing, $c_i, c_j$ are the concentration of the oligos, and $h_i, h_j$ are the directions of the oligos. In other words, oligos anneal with equal rates when they are of opposing direction and overlap beyond a minimal overlap parameter. The corresponding dynamics for the individual oligos are therefore:
\begin{equation}
    \frac{dc_i}{dt} = -k_{anneal} \sum_{j} \Theta(o_{ij} - o_{min})(1-\delta_{h_ih_j})c_ic_j.
\end{equation}
In all simulations reported in the main text, $k_{anneal} = 1$ and $o_{min} = 2$. The rate of annealing and additionally the calculation of the overlap $o_{ij}$ is not affected by any differences in the mutation status of oligos $i$ and $j$.

The reactions themselves are integrated using a sparse representation which is built by comparing all possible clockwise-counterclockwise pairs to find oligo pairs which can potentially anneal. These valid reactions are stored in a cache to speed up future lookups. The equations are integrated by the scipy solve\_ivp function using the RK45 integrator with a relative tolerance of $10^{-5}$ and an absolute tolerance of $10^{-6}$ for a total time of 50.

\subsection{Extending}

Following annealing, extension proceeds by modifying the duplexes present in the system via the following reactions: $ij \rightarrow i'j' $, where $i'$ is the oligo that is obtained by fully extending oligo $i$ along oligo $j$, and $j'$ is the oligo that is obtained by fully extending oligo $j$ along oligo $i$. Essentially, each strand acts as both a primer and a template; oligo $i$ primes its own extension off of oligo $j$ while also serving as the template for the extension of oligo $j$. 

More formally, suppose that clockwise oligo $i$ has start and end $(s_i,e_i)$ and is annealed to counterclockwise oligo $j$ with start and end $(s_j,e_j)$; additionally assume $s_j < e_i < e_j$ so its $3^\prime$ is annealed. Then, it will extend to oligo $i'$ with start and end $(s_i,e_j)$.
Similarly, for counterclockwise oligo $j$, if $s_i < s_j < e_i$, then its $3^\prime$ is annealed and it can also extend to oligo $j'$ with start and end $(s_i,e_j)$. Note that just because oligo $i$ extends does not mean oligo $j$ will also extend.
Furthermore, we do not allow oligos to extend past the maximum VCG length $L$; if clockwise oligo $i$ were to extend past length $L$, it is instead set to have start and end $(s_i,s_i-1)$. For counterclockwise oligo $j$ extending past length $L$, we instead set its new start and end to be $(e_j+1,e_j)$

Note that extension does not change the handedness of the oligo, i.e. $h_{i'} = h_i$. However, extension can change the mutation status $m_i$. In particular, if an oligo $i$ does not yet contain the location where the mutant allele lives ($x_{mut}$), but oligo $j$ does contain $x_{mut}$ upstream of the $3^\prime$ end of oligo $i$, then the resulting oligo $i'$ will adopt the mutation of oligo status $j$, i.e. $m_{i'} = m_j$.

In our simulation, we assume that extension is fast compared to the timescale of the entire melt-anneal-extend cycle and therefore modifies concentration according to the following simple rule:

\begin{equation}
    \Delta c_{i'j'} = \sum_{ij \text{ extends to } i'j'} c_{ij},
\end{equation}
where the sum is taken over all duplexes $ij$ which will produce $i'j'$ upon extension. In other words, extension fully converts all of duplex $ij$ into $i'j'$. Hence the update for the concentration of duplex $ij$ is:
\begin{align}
    \Delta c_{ij} &= -c_{ij} ~~~\text{if duplex $ij$ is capable of extension}, \\
    \Delta c_{ij} &= 0 ~~~~~~~~\text{else}.
\end{align}

Once duplexes are extended, we check to see if new oligos have been created. If so, they are added to the oligo dictionary with the appropriate data of $(s_i, e_i, h_i,m_i)$. 

\subsection{Melting}

After extension, melting splits apart all duplexes whose overlaps do not exceed a threshold overlap $o_{max}$. Schematically, the reaction is the reverse of annealing: $ij \rightarrow i ~+~ j$. Similarly to extension, we assume that melting is fast, so that:
\begin{align}
    \Delta c_{ij} &= -c_{ij} ~~~\text{if} ~~o_{ij} < o_{max}, \\
    \Delta c_{ij} &= 0 ~~~~~~~\text{if} ~~o_{ij} \ge o_{max},
\end{align}
where $o_{ij}$ is the overlap of the oligos in duplex $ij$. In our simulations, we set $o_{max} = 55$, approximately at the value expected for our experimental DNA system.
The corresponding changes in the single-stranded oligo concentrations are therefore:
\begin{equation}
    \Delta c_i = \sum_{j \neq i \text{ s.t. } o_{ij} ~<~ o_{max}} c_{ij}~.
\end{equation}

\subsection{Computational Cleanup}

In order to speed up our simulation, we also implemented a cleanup step to remove oligos with low concentrations. After melting, we removed all oligos which had a concentration below $10^{-5}$, unless they were present only in duplexes.

\subsection{Stalling}

Stalling refers to the empirical observation that the time it takes for extension to proceed after the incorporation of a mismatched base pair can be orders of magnitude greater than the time for extension to proceed after the incorporation of the correct base pair. In order to understand the effect of stalling in a VCG setting, we modified the extension step of our simulation for a subset of oligo configurations. More specifically, since our simulation assumes extension always incorporates the correct base pair, the only time stalled configurations arise in our simulation is if a mutant oligo $i^{mut}$ and a wt oligo $j^{wt}$ have annealed to form a duplex $i^{mut}j^{wt}$ where one of the oligos has a mismatched annealed $3^\prime$ end. Such a duplex would experience stalled extension in either one of two cases:
\begin{enumerate}
    \item Case 1: Oligo $i^{mut}$ has its $3^\prime$ end exactly at the mut allele site ($x_{mut}$), and it has annealed its $3^\prime$ end to oligo $j^{wt}$ where $j^{wt}$ has the wt allele; or,
    \item Case 2: Oligo $j^{wt}$ has its $3^\prime$ end exactly at the at the wt allele site, and it has annealed its $3^\prime$ end to oligo $i^{mut}$ where $i^{mut}$ has the mut allele ($x_{mut}$).
\end{enumerate}
Since these cases are symmetric, we will describe the implementation for the case where $i^{mut}$ is the oligo with the mismatched $3^\prime$ end, but the implementation is identical for $j^{wt}$. In general there is a possibility that both $i^{mut}$ and $j^{wt}$ have mismatched $3^\prime$ ends, but since those configurations cannot occur in the simulation where $o_{min} = 2$, we disregard those.

When Case 1 occurs and $i^{mut}$ has a mismatched $3^\prime$ end, we assume two types of products can be formed: (fully-extended) $i^{mut'}j^{wt'}$ or (stalled)
$i^{mut}j^{wt'}$. In the fully-extended product, extension has proceeded normally on both strands, following the normal extension procedure described above. In the stalled product, extension has proceeded normally for $j^{wt}$, allowing it to extend to $j^{wt'}$, but extension has stalled for $i^{mut}$, preventing it from becoming $i^{mut'}$. In a normal extension simulation, all of the duplex $i^{mut}j^{wt}$ concentration would be converted to fully-extended $i^{mut'}j^{wt'}$ (subject to the normal extension rules described above), but here this conversion is only partial, and some concentration goes to the stalled product.

The relative fraction of stalled product to fully-extended product is related to assumptions we make about the distribution of stalling times. For simplicity, here we assume that extension past a mismatched 3' end is a decay process with an exponential distribution and a characteristic timescale $\tau_{stall}$. Therefore, the probability
$p$ that a stalled oligo extends by the end of a thermal cycle of time $\tau_{cycle}$ is a function:
\begin{equation}
\label{eqn:stall_time_relation}
    p(\tau_{stall}, \tau_{cycle}) = 1-e^{-\frac{\tau_{cycle}}{\tau_{stall}}}. 
\end{equation}
At the continuum level at which we simulate our VCG dynamics, this implies that if a duplex $ij$ exists where the 3$^\prime$ end of oligo $i$ has a mismatch from its corresponding base pair on oligo $j$, then only $pc_{ij}$ will extend into $i'j'$, while $(1-p)c_{ij}$ will become the stalled product $ij'$. Therefore, to incorporate stalling effects into our simulation, we modified the extension step of the simulation to split the conversion of stall-configuration duplexes (Case 1 and Case 2) into the two different types of products (fully-extended and stalled).

In Fig.~6B (top), we set $p = 1.00,~3.68\times10^{-1},~4.54\times10^{-5}$. The simulation accepts $p$ directly as input, and the under our exponential distribution assumption we can then use Eq. \ref{eqn:stall_time_relation} to relate $p$ to the ratio $\frac{\tau_{cycle}}{\tau_{stall}}$.

\subsection{Initial oligo pools}

Having described all dynamic aspects of the simulation, we now described the structured initial conditions that we run the simulation from.

All initial oligo pools are derived from a maximally virtual oligo pool with 12 oligos of length 25, which together tile a length 60 consensus sequence with offsets of 5 bp between each oligo. Following our representation of the oligos, these can be denoted by the following set of segments: $[5i,~5i+25\mod 60]_{i=0,~\dots,~11}$. These 12 oligos are accompanied by their reverse complements, making a total of 24 oligos. Their initial concentrations are set to be equal, and at a value of 10.

These 24 oligos are accompanied by one more oligo which is present at a concentration of 1, and contains a mutant allele. The mutant oligo exactly matches one of the non-mutant oligos, but contains a mutant allele at either a 3$^\prime$-proximal, middle, or $5^\prime$-proximal location along its length. More specifically, the 3$^\prime$-proximal condition corresponds to the mutant allele being present 6~nt from the 3$^\prime$ end of the oligo, the middle condition corresponds to the mutant allele being present 13~nt from the 3$^\prime$ end of the oligo, and the $5^\prime$-proximal condition corresponds to the mutant allele being 21~nt from the 3$^\prime$ end of the oligo.

When varying the virtualness of the pool (as in Fig.~4C), the $V=12$ pool is the full one described above, but the $V=6$ pool is filtered by taking only every other oligo of the original 12 (and its reverse complement, i.e. the following set of segments: $[10i,~10i+25\mod 60]_{i=0,~\dots,~5}$. The $V=3$ pool is filtered even further, so that it consists of just the following three oligos: $[0,~25], ~[20,~45], ~[40,~5]$ (plus their reverse complements).

The different simulation results shown in main text Fig.~4--6 all have the same simulation parameters: identical $o_{min}=2$, $o_{max}=55$, time for annealing (50), and rate of annealing (1). However, they differ in terms of their initial oligo pools. Fig.~4B was generated from a simulation initialized with a $V=12$ oligo pool with a middle mutant. Fig.~4C was generated from oligo pools over the three $V$ conditions (12, 6, and 3), but all with a middle mutant. Fig.~4D was generated from oligo pools over the three mutant location conditions (3$^\prime$-proximal, middle, and $5^\prime$-proximal) but all with $V=12$. Fig.~5 data all comes from a simulation with a 3$^\prime$-proximal mutant and a $V=12$ initial pool. Fig.~6 oligo pools are also all $V=12$ pools, but with mutant oligos where the mutant allele is exactly positioned at the 3$^\prime$ end i.e. it is only 1~nt away from the 3$^\prime$ end. This allows us to consider the effect of stalling as described in the previous section.

\subsection{Simulation Metrics}

We now describe the metrics used to track the state of the simulated oligo pool as it dynamically evolves over multiple thermal melt-anneal-extend cycles.

\subsubsection{Allele Amplification}

To track the growth in concentration of a given allele over thermal cycles, we can keep track of the sum of all concentrations for oligos (including those in duplexes) which contain the allele at the end of each individual thermal cycle. Crucially, we also keep track of the concentration of the allele reverse complement in this way.

However, to understand how well the allele has managed to reproduce itself (and its reverse complement), we need to compare the concentration of the allele to the amount it first started with. Therefore, we define an allele amplification metric to be the concentration of the allele (plus its reverse complement) at a given cycle, divided by the concentration of the allele (plus its reverse complement) in the initial oligo pool. This is the metric used in Fig.~6B (top, bottom).

In addition, we also define a region amplification metric. This metric mimics the effect of measuring allele concentrations in qPCR as is done in the experiments. There, qPCR requires flanking primers for the allele, and hence the region that qPCR measures is not strictly the allele itself, but the allele plus a flanking region. For the mutant, this is easy to define in our simulations, as we choose the flanking region to be equal to the oligo that the mutant initially is on, with an offset of 1~nt on each side. For example, if the mutant initially appeared on oligo $[15,~40]$, then the flanking region would be defined as $[16,~39]$. The wt flanking region would be given by exactly the same region. The allele's regions have to be fully contained within an oligo in order for the oligo's concentration to count to that allele's region's concentration. The region amplification, analogously to the allele amplification, is defined as the concentration of the region (plus its reverse complement) at a given cycle, divided by its initial concentration (plus its reverse complement). This is the metric reported in Fig.~4C--D.

Finally, we define a wt advantage metric in order to quantify the extent to which the wt allele out-competes the mutant allele due to cooperative replication. To compute the wt advantage, we compute the region amplification of the wt after 10 thermal cycles, and the region amplification of the mut after 10 thermal cycles. We then divide the wt region amplification by the mut region amplification, and this ratio gives us the wt advantage. This is the metric reported in Fig.~4C--D.

\subsubsection{Productive Pairing Fraction}

The productive pairing fraction $f^A$ quantifies the average extent to which an allele can form productive duplexes which will template off of it and create more of its reverse complement. In order to compute $f^A$, we first compute $F^A_i$, the productivity factor of an allele $A$ on a (single-stranded) oligo $i$. To compute $F^A_i$, we first identify all single-stranded oligos which can bind to oligo $i$. We then identify a subset of those binding partners which bind with their 3$^\prime$ ends downstream of $A$. We then filter these oligos further to find oligos which don't already overlap the position of allele $A$. This filtered oligo subset is termed the set of productive binding partners since upon extension, they can pick up the reverse complement of oligo $A$. $F^A_i$ is then set to be the fraction of the productive binding partner concentration divided by the total concentration of binding partners. Finally, to compute $f^A$, all $F^A_i$ are summed over the set of all single-stranded oligos $i$ which contain $A$, weighted by their concentrations $c_i$, and divided by the total concentration of both oligos and duplexes which contain $A$. This is the metric which is reported in Fig.~5C(right).

\section{Supplementary Materials}
\label{sec:SupMat}

\subsection{VCG Sequence}\label{supp:VCG_sequence}

\begin{table}[H]
\centering
\caption{Sequences of VCG and mutant oligos used in this study. A1--B12 and B1--B12 define the double-stranded VCG architecture, mapping a 60 bp sequence: \textbf{GCC TTG CGT AAT CTC CAC CTG ACG ACT ATC ATA CAC TGG TCT GTT GTG CTC TAA ATG TCC} in the A-strand orientation.}

\setlength{\tabcolsep}{12pt}
\begin{tabular}{ll}
\toprule
Strand Name & Sequence \\
\midrule
A1 & GCC TTG CGT AAT CTC CAC CTG ACG A \\
A2 & GCG TAA TCT CCA CCT GAC GAC TAT C \\
A3 & ATC TCC ACC TGA CGA CTA TCA TAC A \\
A4 & CAC CTG ACG ACT ATC ATA CAC TGG T \\
A5 & GAC GAC TAT CAT ACA CTG GTC TGT T \\
A6 & CTA TCA TAC ACT GGT CTG TTG TGC T \\
A7 & ATA CAC TGG TCT GTT GTG CTC TAA A \\
A8 & CTG GTC TGT TGT GCT CTA AAT GTC C \\
A9 & CTG TTG TGC TCT AAA TGT CCG CCT T \\
A10 & GTG CTC TAA ATG TCC GCC TTG CGT A \\
A11 & CTA AAT GTC CGC CTT GCG TAA TCT C \\
A12 & TGT CCG CCT TGC GTA ATC TCC ACC T \\
B1 & TCG TCA GGT GGA GAT TAC GCA AGG C \\
B2 & GAT AGT CGT CAG GTG GAG ATT ACG C \\
B3 & TGT ATG ATA GTC GTC AGG TGG AGA T \\
B4 & ACC AGT GTA TGA TAG TCG TCA GGT G \\
B5 & AAC AGA CCA GTG TAT GAT AGT CGT C \\
B6 & AGC ACA ACA GAC CAG TGT ATG ATA G \\
B7 & TTT AGA GCA CAA CAG ACC AGT GTA T \\
B8 & GGA CAT TTA GAG CAC AAC AGA CCA G \\
B9 & AAG GCG GAC ATT TAG AGC ACA ACA G \\
B10 & TAC GCA AGG CGG ACA TTT AGA GCA C \\
B11 & GAG ATT ACG CAA GGC GGA CAT TTA G \\
B12 & AGG TGG AGA TTA CGC AAG GCG GAC A \\
$\text{A1}^{mut}$ & GCC TTG CGT AAT CGC TTC CTG ACG A \\
$\text{A1}^{mut}_{3^\prime-end}$ & GCC TTG CGT AAT CTC CAC GGT AGG A \\
$\text{A1}^{mut}_{5^\prime-end}$ & GCC ATC GGT AAT CTC CAC CTG ACG A \\
\bottomrule
\end{tabular}
\label{tab:oligo_sequences}
\end{table}

\begin{table}[H]
\centering
\caption{Oligo components of VCG mixtures for different virtualness. Sequence of each component is included in Supp.~Sec.~\ref{sec:SupMat} Table \ref{tab:oligo_sequences}}.
\setlength{\tabcolsep}{12pt}
\begin{tabular}{ll}
\toprule
Mixture Name & Components \\
\midrule
VCG12 & A1--B12, B1--B12\\
VCG6 & A1, A3, A5, A7, A9, A11, B1, B3, B5, B7, B9, B11 \\
VCG3 & A1, A5, A9, B1, B5, B9\\
\bottomrule
\end{tabular}
\label{tab:def_virtualness}
\end{table}

\subsection{Primer Sequences}\label{supp:primer_sequences}

\setlength{\tabcolsep}{18pt}
\begin{table}[ht]
\centering
\caption{Primer sequences used for qPCR detection of wildtype and mutant oligos.}
\begin{tabular}{lll}
\toprule
Primer Name & Sequence ($5^\prime \rightarrow 3^\prime$) & Melting Temperature \\
\midrule
\texttt{A1-fwd}     &  GCC TTG CGT AA & 54 \\
\texttt{A1-rev}    &  TCG TCA GGT GGA G & 58\\
\texttt{A1-mut-fwd}  &  GCC TTG CGT AA & 54 \\
\texttt{A1-mut-rev}  &  TCG TCA GGA AGC G & 58\\
\texttt{A1-mut(3$^\prime$)-fwd} & GCC TTG CGT AA & 54\\
\texttt{A1-mut(3$^\prime$)-rev} & TCC TAC CGT G & 52 \\
\texttt{A1-mut(5$^\prime$)-fwd} & GCC ATC GGT AA & 54\\
\texttt{A1-mut(5$^\prime$)-rev} & TCG TCA GGT GGA G & 583E \\
\bottomrule
\end{tabular}
\label{tab:primer_sequences}
\end{table}